\documentclass[draftcls,onecolumn,12pt]{IEEEtran}
\usepackage{color}
\usepackage{graphicx}
\usepackage{epstopdf}
\usepackage{amsmath}
\usepackage{amssymb}
\usepackage[english]{babel}
\usepackage{cite}
\usepackage{rotfloat}
\usepackage{mathtools}
\usepackage[justification=centering]{caption}
\usepackage{breqn}
\usepackage{amsmath}
\usepackage{makecell}
\usepackage{algorithm,algorithmic}
\usepackage{multirow}
\usepackage{subfigure}
\usepackage{booktabs}
\usepackage{colortbl}

\definecolor{kugray5}{RGB}{224,224,224}

\usepackage[normalem]{ulem}
\newcommand\rsout{\bgroup\markoverwith
	{\textcolor{red}{\rule[0.5ex]{2pt}{0.8pt}}}\ULon}



\makeatletter
\newcommand{\ALOOP}[1]{\ALC@it\algorithmicloop\ #1%
	\begin{ALC@loop}}
	\newcommand{\ENDALOOP}{\end{ALC@loop}\ALC@it\algorithmicendloop}

\makeatother

\usepackage{etoolbox}
\let\mybibitem\bibitem
\renewcommand{\bibitem}[1]{%
	\ifstrequal{#1}{nature}
	{\color{blue}\mybibitem{#1}}
	{\color{black}\mybibitem{#1}}%
}

\graphicspath{ {Figures/} }

\newtheorem{remark}{Remark}

\DeclareCaptionLabelSeparator{periodspace}{.\quad}

\renewcommand{\figurename}{Fig.}
\addto\captionsenglish{\renewcommand{\figurename}{Fig.}}
\newcommand\numberthis{\addtocounter{equation}{1}\tag{\theequation}}

\newcommand{\norm}[1]{\left\lVert#1\right\rVert} 
\newcommand{\eq}[1]{\begin{align*}#1\end{align*}} 
\newcommand{\eqn}[1]{\begin{align}#1\end{align}} 
\newcommand{\nt}[1]{\left(#1\right)} 
\newcommand{\nv}[1]{\left[#1\right]} 
\newcommand{\nn}[1]{\left\{#1\right\}} 
\newcommand{\abs}[1]{\left|#1\right|} 

\newcommand{\re}[1]{\mathfrak{R}{\left(#1\right)}}
\newcommand{\im}[1]{\mathfrak{I}{\left(#1\right)}}

\newcommand{\mean}[1]{\mathbb{E} \left\{#1\right\}}

\newcommand{\mH}{\textbf{\textit{H}}}

\newcommand{\mW}{\textbf{\textit{W}}}
\newcommand{\mP}{\textbf{\textit{P}}}

\newcommand{\setR}{\mathbb{R}}
\newcommand{\setA}{\mathcal{A}}

\newcommand{\setAtld}{\tilde{\setA}^{N_t}}
\newcommand{\setN}{\mathcal{N}(\vc)}

\newcommand{\ltabu}{\mathcal{L}}

\newcommand{\vxb}{\textbf{\textit{x}}^{\star}}
\newcommand{\vc}{\textbf{\textit{c}}}

\newcommand{\ve}{\textbf{\textit{e}}} 
\newcommand{\vs}{\textbf{\textit{s}}}
\newcommand{\vx}{\textbf{\textit{x}}}
\newcommand{\vy}{\textbf{\textit{y}}}

\newcommand{\vv}{\textbf{\textit{v}}}
\newcommand{\vn}{\textbf{\textit{n}}}
\newcommand{\vu}{\textbf{\textit{u}}}
\newcommand{\vz}{\textbf{\textit{z}}} 
\newcommand{\vh}{\textbf{\textit{h}}} 

\newcommand{\vq}{\textbf{\textit{q}}}
\newcommand{\vb}{\textbf{\textit{b}}}
\newcommand{\vw}{\textbf{\textit{w}}}

\newcommand{\smt}{\sigma_t^2} 
\newcommand{\smn}{\sigma_n^2} 
\usepackage{setspace}

\begin{document}
	\title{Deep Learning-Aided Tabu Search Detection for Large MIMO Systems}
	\author{Nhan~Thanh Nguyen~
		and~Kyungchun~Lee,~\IEEEmembership{Senior Member,~IEEE}
		\thanks{N. T. Nguyen is with the Department of Electrical and Information Engineering, Seoul National University of Science and Technology,  Seoul 01811, Republic of Korea (e-mail: nhan.nguyen@seoultech.ac.kr).}
		\thanks{K. Lee is with the Department of Electrical and Information Engineering and the Research Center for Electrical and Information Technology, Seoul National University of Science and Technology,  Seoul 01811, Republic of Korea (e-mail: kclee@seoultech.ac.kr).}
		
	}
	
	\maketitle
	
	\begin{abstract}
In this study, we consider the application of deep learning (DL) to tabu search (TS) detection in large multiple-input multiple-output (MIMO) systems. First, we propose a deep neural network architecture for symbol detection, termed the fast-convergence sparsely connected detection network (FS-Net), which is obtained by optimizing the prior detection networks called DetNet and ScNet. Then, we propose the DL-aided TS algorithm, in which the initial solution is approximated by the proposed FS-Net. Furthermore, in this algorithm, an adaptive early termination algorithm and a modified searching process are performed based on the predicted approximation error, which is determined from the FS-Net-based initial solution, so that the optimal solution can be reached earlier. The simulation results show that the proposed algorithm achieves approximately $90\%$ complexity reduction for a $32 \times 32$ MIMO system with QPSK with respect to the existing TS algorithms, while maintaining almost the same performance.

 
	\end{abstract}
	
	\begin{IEEEkeywords}
		MIMO, deep learning, deep neural network, tabu search.
	\end{IEEEkeywords}
	\IEEEpeerreviewmaketitle
	
	\section{Introduction}
	In mobile communications, a large multiple-input multiple-output (MIMO) system is a potential technique to dramatically improve the system’s spectral and power efficiency \cite{ngo2013energy}, \cite{marzetta2010noncooperative}. However, in order for the promised benefits of large MIMO systems to be reaped, significantly increased computational complexity requirements are presented at the receiver when compared to those of the conventional MIMO system \cite{wu2014large, nguyen2019qr}. Therefore, low-complexity near-optimal detection is an important challenge in realizing large MIMO systems \cite{rusek2013scaling, chockalingam2014large, mandloi2017low}. Two major lines of studies have been conducted recently to fulfill that challenge, including proposals of low-complexity near-optimal detection algorithms \cite{vardhan2008low, mohammed2009high, qin2016near, mandloi2017error, som2010improved, mohammed2009low, datta2013novel, hansen2009near, mandloi2017layered, narasimhan2014channel, vsvavc2013soft, nguyen2019qr} and resorting to deep-learning (DL) techniques for symbol detection in massive MIMO systems \cite{farsad2017detection, ye2017power,mohammadkarimi2018deep,samuel2019learning,samuel2017deep,gao2018sparsely}.
	
	\subsection{Recent works}
	Various algorithms for large-MIMO detection have been introduced \cite{vardhan2008low, mohammed2009high, qin2016near, mandloi2017error, som2010improved, mohammed2009low, datta2013novel, hansen2009near, mandloi2017layered, narasimhan2014channel, vsvavc2013soft, nguyen2019qr}. Among them, the tabu search (TS) detector is considered as a complexity-efficient scheme for symbol detection in large MIMO systems. It has been shown that the TS detection algorithm can perform very close to the maximum-likelihood (ML) bound with far lower complexity compared to sphere decoding (SD) and fixed-complexity SD (FSD) schemes in large MIMO systems \cite{rusek2013scaling}, \cite{srinidhi2011layered}. In \cite{srinidhi2009low}, an approach based on reactive TS (RTS) is proposed for near-ML decoding of non-orthogonal $64 \times 64$ space-time block codes (STBCs) with 4-QAM. However, its performance is far from optimal for higher-order QAMs, such as 16- and 64-QAM \cite{srinidhi2009near}. The work in \cite{srinidhi2011layered} proposes an algorithm called layered TS (LTS). This algorithm improves the performance of the TS detection in terms of the bit-error rate (BER) for higher-order QAM in large MIMO systems. However, to achieve a BER of $10^{-2}$ in $32 \times 32$ and $64 \times 64$ MIMO systems with 16-QAM, higher complexities are required than in conventional TS. The random-restart reactive TS (R3TS) algorithm, which runs multiple RTS and chooses the best among the resulting solution vectors, is presented in \cite{datta2010random}. It achieves improved BER performance at the expense of increased complexity. The complexity of R3TS is generally higher than that of RTS to achieve a BER of $10^{-2}$, especially for large antenna configurations and high-order QAMs, such as $64 \times 64$ MIMO with 64-QAM. The work of \cite{zhao2007tabu} has been conducted to further improve TS in terms of complexity, which is based on a reduced number of examined neighbors and an early-termination (ET) criterion. However, it comes at the cost of performance loss; for example, to achieve BER $=10^{-3}$ for $4 \times 4$ MIMO with 16-QAM modulation, the TS algorithm with ET has a 3-dB signal-to-noise ratio (SNR) loss compared to the original TS \cite{zhao2007tabu}. In \cite{nguyen2019qr}, the QR-decomposition-aided TS (QR-TS) algorithm is proposed for achieving considerable complexity reduction without any performance loss.
	
	On the other hand, the application of deep learning (DL) to symbol detection in MIMO systems has recently gained much attention \cite{farsad2017detection, ye2017power,samuel2019learning,samuel2017deep,gao2018sparsely,mohammadkarimi2018deep}. In \cite{farsad2017detection}, three detection algorithms based on deep neural networks (DNNs) are proposed for molecular communication systems, which are shown to perform much better than the prior simple detectors. In contrast, the application of DL to symbol detection in orthogonal frequency-division multiplexing (OFDM) systems are considered in \cite{ye2017power}. Specifically, Ye et al. in \cite{ye2017power} show that the detection scheme based on DL can address channel distortion and detect the transmitted symbols with performance comparable to that of the minimum mean-square error (MMSE) receiver. In \cite{mohammadkarimi2018deep}, the DL-based SD scheme is proposed. In particular, the DL-based SD with the radius of the decoding hypersphere learned by a DNN achieves significant complexity reduction with respect to the conventional SD with a marginal performance loss. In particular, the works of \cite{samuel2019learning} and \cite{samuel2017deep} focus on the design of DNNs for symbol detection in large MIMO systems. Specifically, Samuel et al. in \cite{samuel2019learning} and \cite{samuel2017deep} first investigate the fully connected DNN (FC-DNN) architecture for symbol detection and show that although it performs well for fixed channels, its BER performance is very poor for varying channels. To overcome this problem, a DNN that works for both fixed and varying channels, called the detection network (DetNet), is introduced \cite{samuel2019learning, samuel2017deep}. However, the DetNet requires high computational complexity because of its complicated network architecture, motivating the proposal of the sparsely connected network (ScNet) in \cite{gao2018sparsely} to improve performance and reduce complexity.
	
	\subsection{Contributions}
	
	Although TS detection is considered an efficient symbol-detection algorithm for large MIMO systems \cite{nguyen2019qr, srinidhi2011layered}, it requires many searching iterations to find the optimal solution, causing high computational complexity. The TS algorithm introduced in \cite{zhao2007tabu} uses an ET criterion to terminate the iterative searching process early after a certain number of iterations when no better solution is found. Although this scheme provides complexity reduction, it can result in significant performance loss because the early terminated searching process does not guarantee the optimal solution. However, the number of searching iterations in the TS algorithm can be reduced with only marginal performance loss if a good initial solution and efficient searching/ET strategies are employed, which can be facilitated by DL. More specifically, we found that the initial solution obtained by a DNN is remarkably more reliable than the conventional linear zero-forcing (ZF)/MMSE and ordered successive interference-cancellation (OSIC) solutions. Furthermore, unlike in the cases of the ZF, MMSE, and OSIC receivers, the initial solution generated by an appropriate activation function in the DNN often has signals very close to or exactly the same as the constellation symbols, even before a quantization is applied. This property can be exploited to efficiently determine the reliable/unreliable detected symbols in the initial solution. Based on these aspects, the DL-aided TS algorithm is proposed for complexity reduction of the TS algorithm with ET. Our main contributions are summarized as follows:
	
	\begin{itemize}
		
		\item First, we further optimize the DetNet \cite{samuel2019learning, samuel2017deep} and ScNet \cite{gao2018sparsely} architectures to develop the fast-convergence sparsely connected detection network (FS-Net). Our simulation results show that the proposed FS-Net architecture achieves improved performance and reduced complexity with respect to DetNet and ScNet. As a result, the FS-Net-based solution is taken as the initial solution of the TS algorithm.

		\item In each iteration of the conventional TS algorithm, the move from the current candidate to its best neighbor is made, even when it does not result in a better solution. Therefore, it is possible that no better solution is found after a large number of iterations, but high complexity is required. This motivates us to improve the iterative searching phase of the TS algorithm. Specifically, by predicting the incorrect symbols in the FS-Net-based initial solution, more efficient moves can be made so that the optimal solution is more likely to be reached earlier.
		
		\item For further optimization, we consider the ET criterion incorporated with the FS-Net-based initial solution. In particular, unlike the conventional ET criterion, we propose using an adaptive cutoff factor, which is adjusted based on the accuracy of the FS-Net-based initial solution. As a result, when the initial solution is likely to be accurate, a small number of searching iterations is taken, which leads to a reduction in the overall complexity of the TS algorithm.
		
	\end{itemize}
	
	The rest of the paper is organized as follows: Section II presents the system model. Section III reviews and analyzes the complexity of the prior DNNs architectures for symbol detection, namely, the FC-DNN, DetNet, and ScNet, followed by the proposal of the FS-Net architecture. Section IV presents the DL-aided TS detection algorithm. In Section V, the simulation results are shown. Finally, the conclusions are presented in Section VI.

	\textit{Notations}: Throughout this paper, scalars, vectors, and matrices are denoted by lower-case, bold-face lower-case, and bold-face upper-case letters, respectively. The $(i,j)$th element of a matrix $\textbf{\textit{A}}$ is denoted by $a_{i,j}$, whereas $(\cdot)^T$ and $(\cdot)^H$ denote the transpose and conjugate transpose of a vector, respectively. Furthermore, $\abs{\cdot}$ and $\norm{\cdot}$ represent the absolute value of a scalar and the norm of a vector or matrix, respectively. The expectation operator is denoted by $\mean{\cdot}$, whereas $\sim$ means $\textit{distributed as}$.
	
	\section{System Model}

	We consider the uplink of a multi-user MIMO system with $N_r$ receive antennas, where the total number of transmit antennas among all users is $N_t$. The received signal vector $\tilde{\vy}$ is given by
	\eqn{
		\label{complex SM}
		\tilde{\vy} = \tilde{\mH} \tilde{\vs} + \tilde{\vn},
	}
	where $\tilde{\vs} = \nv{\tilde{s}_1, \tilde{s}_2, \ldots, \tilde{s}_{N_t}}^T$ is the vector of transmitted symbols. We assume that $\mean{\abs{\tilde{s}_i}^2} = \smt$, where $\smt$ is the average symbol power, and $\tilde{\vn}$ is a vector of independent and identically distributed (i.i.d.) additive white Gaussian noise (AWGN) samples, $\tilde{n}_i \sim \mathcal{CN}(0,\smn)$. Furthermore, $\tilde{\mH}$ denotes an $N_r \times N_t$ channel matrix consisting of entries $\tilde{h}_{i,j}$, where $\tilde{h}_{i,j}$ represents the complex channel gain between the $j$th transmit antenna and the $i$th receive antenna. The transmitted symbols $\tilde{s}_i, i = 1, 2, \ldots, N_t,$ are independently drawn from a complex constellation $\tilde{\setA}$ of $\tilde{Q}$ points. The set of all possible transmitted vectors forms an $N_t$-dimensional complex constellation $\setAtld$ consisting of $\tilde{Q}^{N_t}$ vectors, i.e., $\tilde{\vs} \in \setAtld$.
	
	The complex signal model \eqref{complex SM} can be converted to an equivalent real signal model
	\eqn{
		\vy = \mH \vs + \vn, \label{real SM}
	}
	where $\vs, \vy, \vn,$ and $\mH$ given by
	\eq{
		\begin{bmatrix}
			\re{\tilde{\vs}}\\
			\im {\tilde{\vs}}
		\end{bmatrix}, 
		\begin{bmatrix}
			\re{\tilde{\vy}}\\
			\im {\tilde{\vy}}
		\end{bmatrix},
		\begin{bmatrix}
			\re{\tilde{\vn}}\\
			\im {\tilde{\vn}}
		\end{bmatrix}, \text{ and }
		\begin{bmatrix}
			\re {\tilde{\mH}}  &-\im {\tilde{\mH}}\\
			\im {\tilde{\mH}}  &\re {\tilde{\mH}}
		\end{bmatrix},
	}
	respectively denote the $\nt{N \times 1}$-equivalent real transmitted signal vector, $\nt{M \times 1}$-equivalent real received signal, AWGN noise signal vectors, and $(M \times N)$-equivalent real channel matrix, with $N = 2N_t, M = 2N_r$. Here, $\re {\cdot}$ and $\im {\cdot}$ denote the real and imaginary parts of a complex vector or matrix, respectively. Then, the set of all possible real-valued transmitted vectors forms an $N$-dimensional constellation $\setA^N$ consisting of $Q^N$ vectors, i.e., $\vs \in \setA^N$. In this work, we use the equivalent real-valued signal model in \eqref{real SM} because it can be employed for both the TS algorithm and DNNs.
	
	\subsubsection{Conventional optimal solution}
	The ML solution can be written as
	\eqn {
		\hat{\vs}_{ML} = \arg \min_{\vs \in \setA^{N}} \phi(\vs), \label{ML_solution}
	}
	where $\phi(\vs) = \norm {\vy - \mH \vs}^2$ is the ML metric of $\vs$. The computational complexity of ML detection in \eqref{ML_solution} is exponential with $N$ \cite{srinidhi2011layered}, which results in extremely high complexity for large MIMO systems, where $N$ is very large. 
	
	\subsubsection{DNN-based solution}
	A DNN can be modeled and trained to approximate the transmitted signal vector $\vs$. The solution obtained by a DNN with $L$ layers can be formulated as
	\begin{align*}
		\hat{\vs} = \mathcal{Q} \left( \hat{\vs}^{[L]} \right),
	\end{align*}
	where $\mathcal{Q} (\cdot)$ is the element-wise quantization operator that quantizes $ \hat{s}^{[L]}_n \in \setR$ to $\hat{s}_n \in \setA, n=1, \ldots,N$. Here, $\hat{\vs}^{[L]}$ is the output vector at the $L$th layer, which can be expressed as
	\begin{align*}
		\hat{\vs}^{[L]} = f^{[L]} \left( f^{[L-1]} \left( \ldots \left(f^{[1]} \left(\vx^{[1]}; \mP^{[1]} \right); \ldots \right); \mP^{[L-1]} \right); \mP_L \right), \numberthis \label{DNN_1}
	\end{align*}
	where
	\begin{align*}
		f^{[l]} \left(\vx^{[l]}; \mP^{[l]} \right) = \sigma^{[l]} \left( \mW^{[l]} \vx^{[l]} + \vb^{[l]} \right) \numberthis \label{f^{[l]}}
	\end{align*}
	represents the nonlinear transformation in the $l$th layer with the input vector $\vx^{[l]}$, the activation function $\sigma^{[l]}$, and $\mP^{[l]} = \left\{\mW^{[l]}, \vb^{[l]}\right\}$ consisting of the weighting matrix $\mW^{[l]}$ and bias vector $\vb^{[l]}$. We see that \eqref{DNN_1} indicates the serial nonlinear transformations in the DNN that maps the input $\vx^{[1]}$, including the information contained in $\vy$ and $\mH$, to the output $\hat{\vs}^{[L]}$.
	
	In large MIMO systems, many hidden layers and neurons are required for the DNN to extract meaningful features and patterns from the large amount of input data to provide high accuracy. Furthermore, the high-dimension signals and large channel matrix lead to the large input vector $\vx$, which requires large $\mW^{[l]}$ and $\vb^{[l]}$ for the transformation in \eqref{f^{[l]}}. As a result, the computational complexity of the detection network typically becomes very high in large MIMO systems. 
	
	\section{DNNs for MIMO Detection}
	
	In this section, we first analyze the architecture designs and complexities of three existing DNNs for MIMO detection in the literature, namely, FC-DNN \cite{samuel2019learning}, DetNet \cite{samuel2019learning}, and ScNet \cite{gao2018sparsely}. This motivates us to further optimize them and propose a novel DNN for performance improvement and complexity reduction in MIMO detection.
	
	\subsection{FC-DNN, DetNet, and ScNet architectures}
	
	\subsubsection{FC-DNN architecture}
	
	The application of the well-known FC-DNN architecture for MIMO detection is investigated in \cite{samuel2019learning, samuel2017deep}. In this FC-DNN architecture, the input vector contains all the received signal and channel entries, i.e., $\vx = \{\vy, \mH\}$. The performance of the FC-DNN is examined in two scenarios: fixed and varying channels. It is shown in \cite{samuel2019learning} that the FC-DNN architecture performs well for fixed channels; however, this is an impractical assumption. In contrast, for varying channels, its performance is very poor. Therefore, it cannot be employed for symbol detection in practical MIMO systems, and a more sophisticated DNN architecture is required for this purpose. 
	
	\subsubsection{DetNet}
	\begin{figure}[t]
		\centering
		\includegraphics[scale=0.8]{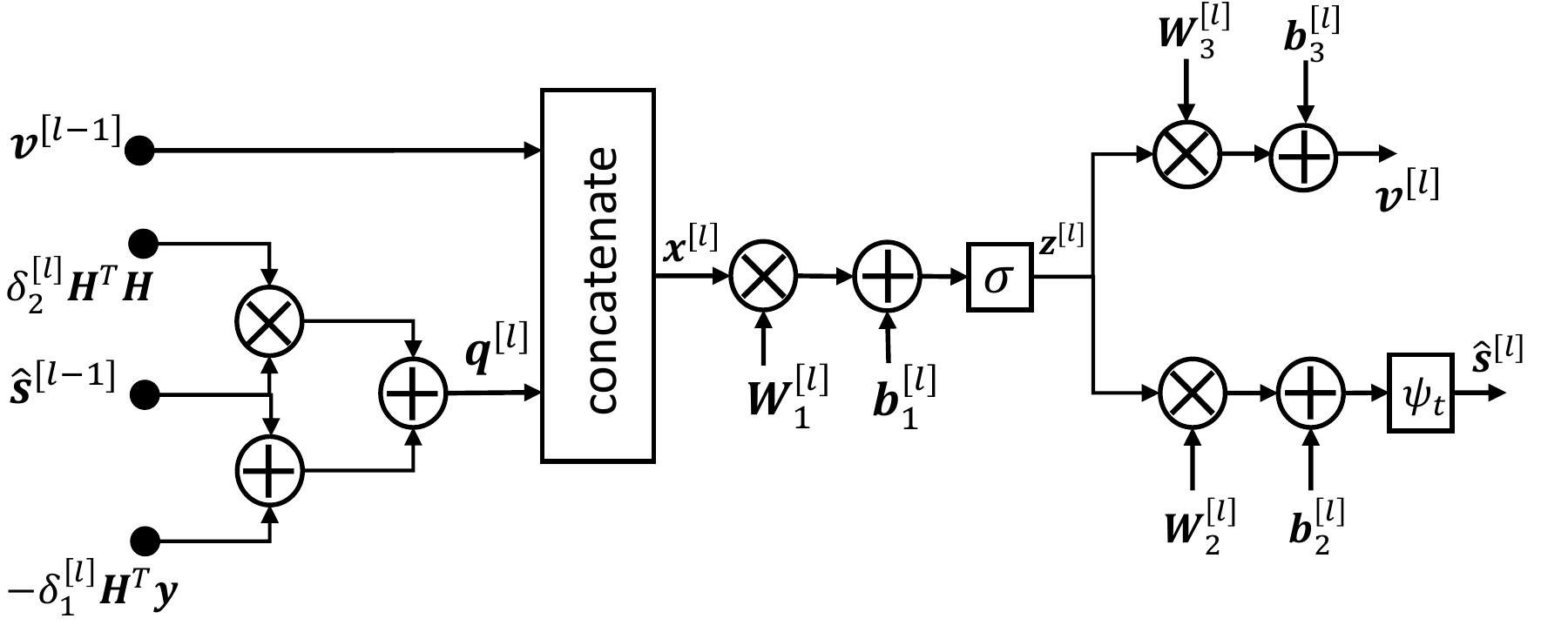}
		\caption{The $l$th layer of the DetNet architecture}
		\label{fig_detnet}
	\end{figure}
	
	In the DetNet, the transmitted signal vector is updated over $L$ iterations corresponding to $L$ layers of the neural network based on mimicking a projected gradient descent-like ML optimization, which leads to iterations of the form \cite{samuel2019learning, samuel2017deep}
	\begin{align*}
		\hat{\vs}^{[l+1]} 
		&= \Pi \left[ \vs - \delta^{[l]} \frac{\partial \norm {\vy - \mH \vs}^2}{\partial \vs} \right]_{\vs = \hat{\vs}^{[l]}} \\
		&= \Pi \left[ \hat{\vs}^{[l]} - \delta^{[l]} \mH^T \vy + \delta^{[l]} \mH^T \mH \hat{\vs}^{[l]} \right], \numberthis \label{detnet_1}
	\end{align*}
	where $\Pi[\cdot]$ denotes a nonlinear projection operator and $\delta^{[l]}$ is a step size. 
	
\begin{figure}[t]
		\centering
		\includegraphics[scale=0.6]{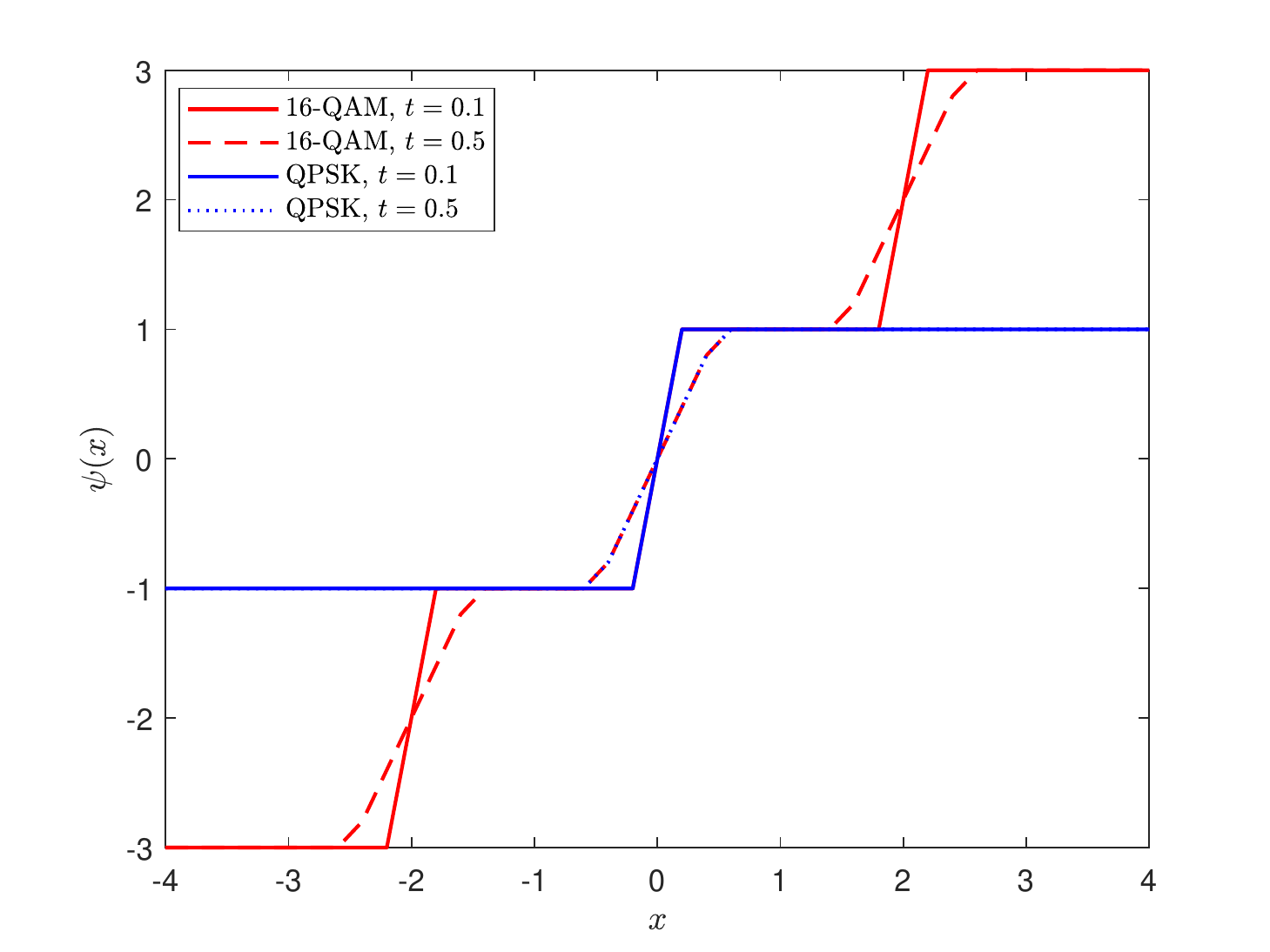}
		\caption{$\psi_t(x)$ in DetNet, ScNet, and FS-Net.}
		\label{fig_phi}
	\end{figure}

	The operation and architecture of the $l$th layer of the DetNet is illustrated in Fig. \ref{fig_detnet}. It is shown that $\hat{\vs}^{[l]}$ and $\vv^{[l]}$, which are not only the output of the $l$th layer but also the input of the $(l+1)$th layer, are updated as follows:
	\begin{align*}
		\vq^{[l]} &= \hat{\vs}^{[l-1]} - \delta_1^{[l]} \mH^T \vy + \delta_2^{[l]} \mH^T \mH \hat{\vs}^{[l-1]}, \numberthis \label{d_q}\\
		\vx^{[l]} &= \left[\vv^{[l-1]}, \vq^{[l]} \right]^T, \numberthis \label{d_x}\\
		\vz^{[l]} &= \sigma \left(\mW_1^{[l]} \vx^{[l]} + \vb_1^{[l]} \right), \numberthis \label{d_z}\\
		\hat{\vs}^{[l]} &= \psi_t \left(\mW_2^{[l]} \vz^{[l]} + \vb_2^{[l]}\right), \numberthis \label{d_s}\\
		\vv^{[l]} &= \mW_3^{[l]} \vz^{[l]} + \vb_3^{[l]}, \numberthis \label{d_v}
	\end{align*}
	where $\hat{\vs}^{[0]}$ and $\vv^{[0]}$ are the input of the first layer $(l=1)$ of the network, which are initialized as $\hat{\vs}^{[0]} = \vv^{[0]} = \textbf{0}$, with $\textbf{0}$ being an all-zero vector of an appropriate size. In \eqref{d_x}, $\vq^{[l]}$ and $\vv^{[l-1]}$ are concatenated into a single input vector $\vx^{[l]}$. In \eqref{d_z}, $\sigma(\cdot)$ is the rectified linear unit (ReLU) activation function. Furthermore, $\psi(\cdot)$ in \eqref{d_s}, defined as
	\begin{align*}
		\psi_t(x) = -q + \frac{1}{\abs{t}} \sum_{i \in \Omega}[\sigma(x + i + t) - \sigma(x + i- t)]
	\end{align*}
	with $q=1, \Omega = \{0\}$ for QPSK and $q=3, \Omega = \{-2,0,2\}$ for 16-QAM, guarantees that the amplitudes of the elements of $\hat{\vs}^{[l]}$ are in the range $[-1,1]$ for QPSK and $[-3,3]$ for 16-QAM, as illustrated in Fig. \ref{fig_phi}. The final detected symbol vector is given as $\hat{\vs} = \mathcal{Q} \left( \hat{\vs}^{[L]} \right)$, where $\mathcal{Q} \left( \cdot \right)$ quantizes each element of $\hat{\vs}^{[L]}$ to its closest real-constellation symbol in $\setA$.
	
	In the training phase, the weights and biases of the DetNet are optimized by minimizing the loss function
	\begin{align*}
		\mathcal{L} (\vs, \hat{\vs}) = \sum_{l=1}^{L} \log (l) \norm{\vs - \hat{\vs}^{[l]}}^2, \numberthis \label{loss_detnet}
	\end{align*}
	which measures the total weighted distance between the transmitted vector $\vs$ and the outputs of all the layers, i.e., $\hat{\vs}^{[l]}, l=1,\ldots,L$. The DetNet is trained to optimize the parameter set $\left\{ \mW_i^{[l]}, \vb_i^{[l]} \right\}$, $i=1,2,3$, $l=1,\ldots,L$, such that $\mathcal{L} (\vs, \hat{\vs})$ is minimized. As a result, $\hat{\vs}$ can converge to $\vs$.
	
	We now consider the computational complexity of the DetNet, which is defined as the total number of additions and multiplications required in \eqref{d_q}--\eqref{d_v}. The computations of $\mH^T \vy$ and $\mH^T \mH$ require $N(2M-1)$ and $N^2(2M-1)$ operations, respectively, and are performed once in the first layer. The complexities required in \eqref{d_q}, \eqref{d_z}--\eqref{d_v} depend on the modulation scheme, as follows:
	\begin{itemize}
		\item For QPSK, the sizes of $\vv^{[l]}$, $l=1,\ldots,L$, is set to $N \times 1$ \cite{samuel2019learning}, leading to $\vx^{[l]} \in \setR^{2N \times 1}$. As a result, the sizes of $\mW_1^{[l]}, \mW_3^{[l]}, \vb_1^{[l]}$, and $\vb_3^{[l]}$ can be inferred from the size of $\vz^{[l]}$, which is set to $2N \times 1$ \cite{samuel2019learning}, as follows: $\mW_1^{[l]} \in \setR^{2N \times 2N}, \vb_1^{[l]} \in \setR^{2N \times 1}, \mW_3^{[l]} \in \setR^{N \times 2N}$, and $\vb_3^{[l]} \in \setR^{N \times 1}$. Furthermore, because $\hat{\vs}^{[l+1]} \in \setR^{N \times 1}$, we have $\mW_2^{[l]}\in \setR^{N \times 2N}$, and $\vb_2^{[l]} \in \setR^{N \times 1}$. Then, given $\mH^T \vy$ and $\mH^T \mH$, the total complexity required in each layer is $18N^2 + N$ operations, including $2N^2+N$, $8N^2$, $4N^2$, and $4N^2$ operations for \eqref{d_q}, \eqref{d_z}--\eqref{d_v}, respectively. Therefore, the total complexity of all $L$ layers of the DetNet is given as
		\begin{align*}
			\mathcal{C}_{\text{DetNet}}^{\text{QPSK}} 
			&= N(2M-1) + N^2(2M-1) + L(18N^2 + N)\\
			&= (18L+2M-1) N^2 + (2M-1+L)N. \numberthis \label{c_detnet_qpsk}
		\end{align*}
		
		\item For 16-QAM, $\vv^{[l]}$ and $\vz^{[l]}$ are set to $\vv^{[l]} \in \setR^{2N \times 1}$ and $\vz^{[l]} \in \setR^{4N \times 1}$ \cite{samuel2019learning}. Therefore, we have $\vx^{[l]} \in \setR^{3N \times 1}$, resulting in $\mW_1^{[l]} \in \setR^{4N \times 3N}, \vb_1^{[l]} \in \setR^{4N \times 1} , \mW_2^{[l]} \in \setR^{N \times 4N}$, $\vb_2^{[l]} \in \setR^{N \times 1}$, $\mW_3^{[l]}\in \setR^{2N \times 4N}$, and $\vb_3^{[l]} \in \setR^{2N \times 1}$. Then, given $\mH^T \vy$ and $\mH^T \mH$, the complexities required in \eqref{d_q} and \eqref{d_z}--\eqref{d_v} are $2N^2+N$, $24N^2$, $8N^2$, and $16N^2$ operations, respectively. As a result, the total complexity required in each layer of the DetNet with 16-QAM is $50 N^2 + N$ operations. Therefore, the total complexity of all $L$ layers of the DetNet is given as
		\begin{align*}
			\mathcal{C}_{\text{DetNet}} ^{\text{16-QAM}} 
			&= N(2M-1) + N^2(2M-1) + L(50 N^2 + N)\\
			&= (50 L+2M-1) N^2 + (2M-1+L)N. \numberthis \label{c_detnet_qam}
		\end{align*}
	\end{itemize}
	
	It is observed from \eqref{c_detnet_qpsk} and \eqref{c_detnet_qam} that for both QPSK and 16-QAM, the complexity of the DetNet can be substantially high in large MIMO systems, where $N$ is large. This high complexity is due to the use of the additional input vector $\vv$ and the full connections between the input and output vectors in every layer. Furthermore, the complicated architecture of the DetNet makes it difficult to optimize, which results in its relatively low performance, as will be shown in Section \ref{sec_sim_result}. Therefore, the ScNet was introduced in \cite{gao2018sparsely} for complexity reduction and performance improvement.
	
	\subsubsection{ScNet}
	
	The ScNet also follows the update process in \eqref{detnet_1}, but it simplifies the DetNet architecture based on the following observations:
	\begin{itemize}
		\item While $\vq$ contains information of $\mH$, $\vy$, and $\hat{\vs}$, the additional input vector $\vv$ does not contain any other meaningful information. Therefore, $\vv$ is removed in the ScNet architecture. As a result, $\{\mW_3, \vb_3\}$ is also removed. It is shown in \cite{gao2018sparsely} that this simplification leads not only to reduced complexity and training time, but also improved performance.
		
		\item Furthermore, it is observed in \eqref{detnet_1} that the first element of $\hat{\vs}^{[l+1]}$ only depends on the first element of $\vq^{[l]}$, which implies that the full connection between all elements of $\vq^{[l]}$ and $\hat{\vs}^{[l+1]}$ is unnecessary. As a result, the input and output of each layer of the ScNet are directly connected in the element-wise manner. Consequently, $\left\{\mW_2^{[l]}, \vb_2^{[l]}\right\}$ is removed, and the weight matrix $\mW_1^{[l]}$ is reduced to a weight vector $\vw^{[l]}$ of size $3N \times 1$.
	\end{itemize} 
	
\begin{figure}[t]
		\centering
		\includegraphics[scale=1.2]{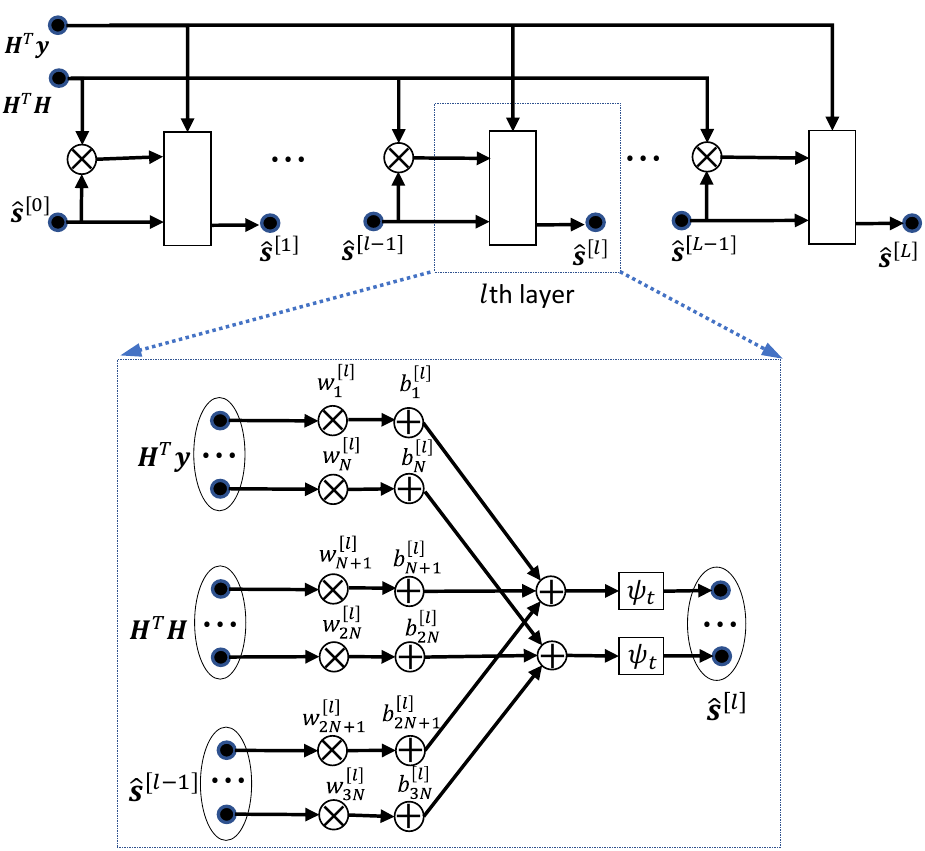}
		\caption{The ScNet architecture}
		\label{fig_scnet}
	\end{figure}

	The operation and architecture of the ScNet is illustrated in Fig. \ref{fig_scnet}. Similar to DetNet, ScNet is initialized with $\hat{\vs}^{[0]} = \vv^{[0]} = \textbf{0}$. Then, the output of the $l$th layer, i.e., $\hat{\vs}^{[l]}$, is updated as follows:
	\begin{align*}
		\vx^{[l]} &= [\mH^T \vy, \mH^T \mH \hat{\vs}^{[l-1]}, \hat{\vs}^{[l-1]}]^T, \numberthis \label{s_x}\\
		\hat{\vs}^{[l]} &= \psi_t (\vw^{[l]} \odot \vx^{[l]} + \vb_{l}), \numberthis \label{s_s}
	\end{align*}
	where $\vw^{[l]} \odot \vx^{[l]}$ denotes the element-wise multiplication of $\vw^{[l]}$ and $\vx^{[l]}$. Given $\mH^T \mH$, the computation of $\mH^T \mH \hat{\vs}^{[l]}$ in \eqref{s_x} requires $2N^2-N$ operations. Furthermore, because $\vx^{[l]} \in \setR^{3N \times 1}$, we have $\vw^{[l]}, \vb^{[l]} \in \setR^{3N \times 1}$, and the computation in \eqref{s_s} requires only $6N$ operations. Consequently, the complexity of each layer of the ScNet architecture is $2N^2 + 5N$. Taking the complexities of computing $\mH^T \vy$ and $\mH^T \mH$ into consideration, the ScNet architecture requires
	\begin{align*}
		\mathcal{C}_{\text{ScNet}} 
		&= N(2M-1) + N^2(2M-1) + L(2N^2 + 5N) \\
		&= (2M-1+2L)N^2 + (2M-1+5L)N \numberthis \label{c_scnet}
	\end{align*}
	operations in total. Compared to the DetNet expressed in \eqref{c_detnet_qpsk} and \eqref{c_detnet_qam}, it is observed that the ScNet requires much lower complexity. Furthermore, the simulation results in \cite{gao2018sparsely} show that for BER $=10^{-4}$, the ScNet achieves an approximate SNR gain of 1 dB over the DetNet. 
	
	However, one drawback of the ScNet is that the input vector $\vx$ has the size of $(3N \times 1)$, which is three times larger than that of $\vq$ in the DetNet for containing the information for $\vy$, $\mH$, and $\hat{\vs}^{[l-1]}$. This may result in unnecessary computational complexity of the ScNet. Furthermore, both the DetNet and ScNet employ the loss function \eqref{loss_detnet} for the optimization of the weights and biases. This loss function is able to minimize the distance between $\vs$ and $\hat{\vs}$, which allows $\hat{\vs}$ to converge to $\vs$ after a certain number of updates, which is equal to the number of layers in the DetNet and ScNet. However, it does not guarantee fast convergence. Meanwhile, if $\hat{\vs}$ converges to $\vs$ faster, a smaller number of layers can be required to achieve the same accuracy. In other words, for the same number of layers, if a better loss function is employed, then the performance can be improved. These observations on the input vector and the loss function of the DetNet and ScNet motivate us to propose the FS-Net for complexity reduction and performance improvement.
	
	\subsection{Proposed FS-Net architecture}
	\subsubsection{Network architecture}
	
	Inheriting the DetNet and ScNet, the proposed FS-Net is also motivated by the updating process in \eqref{detnet_1}. We note that \eqref{detnet_1} can be rewritten as
	\begin{align*}
		\hat{\vs}^{[l+1]} 
		&= \Pi \left[ \hat{\vs}^{[l]}  + \delta^{[l]} (\mH^T \mH \hat{\vs}^{[l]} - \mH^T \vy) \right], \numberthis \label{detnet_2}
	\end{align*}
	which shows that the contributions of $\hat{\vs}^{[l]}$ and $\mH^T \mH \hat{\vs}^{[l]} - \mH^T \vy$ to $\hat{\vs}^{[l+1]}$ are different. Therefore, their elements should be processed by different weights and biases. Furthermore, \eqref{detnet_2} also implies that the elements at the same position of $\mH^T \mH \hat{\vs}^{[l]}$ and $\mH^T \vy$ can be multiplied by the same weight. Therefore, in the proposed FS-Net, we set the input vector of the $(l+1)$th layer to
	\begin{align*}
		\vx^{[l]} = \left[ \hat{\vs}^{[l]}, \mH^T \mH \hat{\vs}^{[l]} - \mH^T \vy \right]^T \in \setR^{2N \times 1},
	\end{align*}
	whose size is only $2/3$ that of $\vx^{[l]}$ in \eqref{s_x} for the ScNet. Furthermore, the FS-Net follows the sparse connection of ScNet. Consequently, in each layer of the FS-Net, there are only $2N$ element-wise connections between the input and output, whereas the ScNet has $3N$. 
	
\begin{figure}[t]
		\centering
		\includegraphics[scale=1.2]{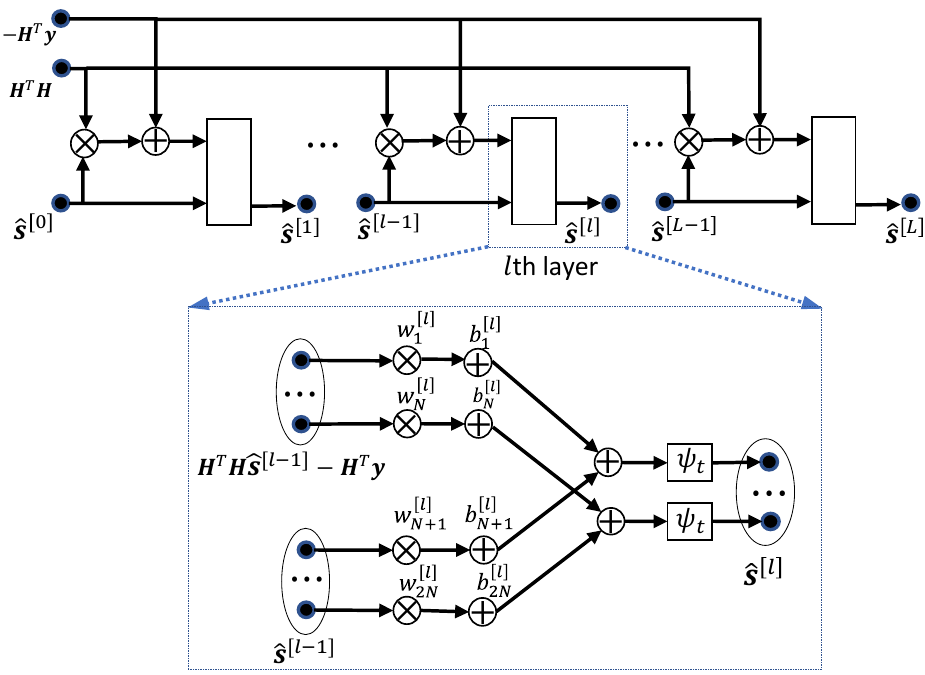}
		\caption{The FS-Net architecture}
		\label{fig_dscnet}
	\end{figure}

	The operation and architecture of the proposed FS-Net network is illustrated in Fig. \ref{fig_dscnet}. Furthermore, Algorithm \ref{al_dscnet} summarizes the FS-Net scheme for MIMO detection. The output of each layer is updated in step 4, where $\vx^{[l]}$ is obtained in step 3 with the requirement of $2N^2$ operations to compute $\mH^T \mH \hat{\vs}^{[l]} - \mH^T \vy$ when $\mH^T \vy$ and $\mH^T \mH$ are given. In the FS-Net architecture, we have $\vw^{[l]}, \vb^{[l]} \in \setR^{2N \times 1}$. Therefore, the computation in each layer requires only $2N^2 + 4N$ operations. We recall that the complexities of computing  $\mH^T \vy$ and $\mH^T \mH$ are $N(2M-1)$ and $N^2(2M-1)$ operations. As a result, the complexity of the entire FS-Net is given as
	\begin{align*}
		\mathcal{C}_{\text{FS-Net}} 
		&= N(2M-1) + N^2(2M-1) + L(2N^2 + 4N )\\
		&= (2M-1+2L)N^2 + (2M-1+4L)N, \numberthis \label{c_dsn}
	\end{align*}
	which is less than that of the ScNet architecture by $LN$ operations and considerably lower than the complexity of the DetNet given in \eqref{c_detnet_qpsk} and \eqref{c_detnet_qam}.
	
	\subsubsection{Loss function}
	We consider the correlation between $\hat{\vs}^{[l]}$ and $\vs$ in the loss function for better training the FS-Net. Specifically, the loss function of the FS-Net is redefined as
	\begin{align*}
		\mathcal{L} (\vs, \hat{\vs}) = \sum_{l=1}^{L} \log (l) \left[\alpha \norm{\vs - \hat{\vs}^{[l]}}^2 + \beta r (\hat{\vs}^{[l]}, \vs) \right], \numberthis \label{loss_dscnet}
	\end{align*} 
	where $r (\hat{\vs}^{[l]}, \vs) = 1 -  \frac{\abs{\vs^T \hat{\vs}^{[l]}}}{\norm{\vs}  \lVert \hat{\vs}^{[l]}\rVert }$. Based on the Cauchy--Schwarz inequality, we have $|\vs^T \hat{\vs}^{[l]}| \leq \norm{\vs} \lVert \hat{\vs}^{[l]}\rVert$, where equality occurs if $ \vs = c \hat{\vs}^{[l]}$ with a constant $c$. Therefore, we have $r(\hat{\vs}^{[l]}, \vs) \geq 0$, and $r(\hat{\vs}^{[l]}, \vs) = 0$ if $\vs = c \hat{\vs}^{[l]}$. 
	
	The DetNet, ScNet, and FS-Net schemes initialize $\hat{\vs}^{[0]}$ as $\textbf{0}$ and update $\hat{\vs}^{[l]}$ over $L$ layers to approximate $\vs$. This can be considered as sequential moves starting from the coordinate $\textbf{0}$ over $\hat{\vs}^{[1]}, \ldots, \hat{\vs}^{[L-1]}$ to reach $\hat{\vs}^{[L]} \approx \vs$. By minimizing the loss function in \eqref{loss_dscnet}, we have $r (\hat{\vs}^{[l]}, \vs) \rightarrow 0$, or equivalently, $\vs \approx c \hat{\vs}^{[l]}, l = 1, 2, \ldots, L$, which enables the moves to be in a specific direction, which is the position of $\vs$ in a hypersphere. This can shorten the path of the moves to reach $\vs$, which results in the reduced number of required layers in the FS-Net. In \eqref{loss_dscnet}, $\alpha$ and $\beta$ are used to adjust the contributions of $\lVert \vs - \hat{\vs}^{[l]}\rVert ^2$ and $r (\hat{\vs}^{[l]}, \vs)$ to the loss function, which are optimized by simulations. Our simulation results in Section V show that the proposed FS-Net achieves not only complexity reduction, but also performance improvement with respect to the conventional DetNet and ScNet schemes.
	
	\begin{algorithm}[t]
		\caption{FS-Net scheme for MIMO detection}
		\label{al_dscnet}
		\begin{algorithmic}[1]
			\REQUIRE $\mH, \vy$.
			\ENSURE $\hat{\vs}$.
			\STATE {$\hat{\vs}^{[0]} = \textbf{0}$}
			\FOR {$l = 1 \rightarrow L$}
			\STATE {$\vx^{[l]} = \left[ \hat{\vs}^{[l-1]}, \mH^T \mH \hat{\vs}^{[l-1]} - \mH^T \vy \right]^T$}
			\STATE {$\hat{\vs}^{[l]} =  \psi_t (\vw^{[l]} \odot \vx^{[l]} + \vb)$}
			\ENDFOR 
			\STATE {$\hat{\vs} = \mathcal{Q} \left( \hat{\vs}^{[L]} \right)$}
		\end{algorithmic}
	\end{algorithm}
	
	\section{Deep Learning-Aided Tabu Search Detection}
	
	\subsection{Problem formulation} 
	The TS algorithm starts with an initial candidate and sequentially moves over $\mathcal{I}_{\text{UB}}$ candidates for $\mathcal{I}_{\text{UB}}$ iterations. In each iteration, all the non-tabu neighbors of the current candidate $\vc$ are examined to find the best neighbor $\vxb$ with the smallest ML metric, i.e., 
	\begin{align*}
		\vxb = \arg \min_{\vx \in \setN}  \phi(\vx), \numberthis \label{best_nb}
	\end{align*}
	where $\setN$ consists of non-tabu neighboring vectors inside alphabet $\setA^{N}$ with the smallest distance to $\vc$, i.e.,
	\eq {
		\mathcal{N}(\vc) = \nn{\vx \in \setA^{N}\backslash {\ltabu}, \abs {\vx - \vc} = \theta_{min}}, \numberthis \label{nb_def}
	}
	where $\setA^{N} \backslash {\ltabu}$ denotes the alphabet $\setA^{N}$ excluding the tabu vectors kept in the tabu list $\ltabu$, and $\theta_{min}$ is the minimum distance between two constellation points in a plane. Furthermore, the ML metric $\phi (\vx)$ can be expressed as \cite{nguyen2019qr}
	\begin{align*}
		\phi (\vx) = \norm {\vu + \vh_d \delta_d}^2, \numberthis \label{phi_x}
	\end{align*}
	where $\vu = \vy - \mH \vc$, $\delta_d$ is the single nonzero element of $\vc - \vx = \nv {0, \ldots, 0, \delta_d, 0, \ldots, 0}^T$ and $\vh_d$ is the $d$th column of $\mH$. In this study, we refer to $d$ as the \textit{difference position} of a neighbor, in which the candidate and its neighbor are different. For example, if the current candidate is $\vc=[1, -3, 1, 3]^T$, then $d=3$ is the difference position for $\vx = [1, -3, -1, 3]^T$ because $\vc $ and $\vx $ are only different with respect for the third element.
	
	After the best neighbor is determined with \eqref{best_nb}, it becomes the candidate in the next iteration, and the determination of the best neighbor of a new candidate is performed. By this iterative manner, the final solution $\hat{\vs}_{TS}$ is determined as the best candidate visited so far, i.e.,
	\begin{align*}
		\hat{\vs}_{TS} = \arg \min_{\vc \in \mathcal{V}} \left\{ \phi (\vc) \right\}
	\end{align*}
	where $\mathcal{V}$ is the set of all visited candidates over $\mathcal{I}_{\text{UB}}$ searching iterations.
	
	The computational complexity of TS algorithms is proportional to the number of searching iterations, i.e., $\mathcal{I}_{\text{UB}}$. In large MIMO systems, the number of neighbors in each iteration and the dimension of the neighboring vectors are large. Therefore, the complexity to find the best neighbor in each iteration becomes high in large MIMO systems. Furthermore, an extremely large $\mathcal{I}_{\text{UB}}$ is required to guarantee that the near-optimal solution is found. Consequently, the complexity of the TS algorithm can be excessively high in large MIMO systems. To reduce the complexity of TS in large MIMO systems, an ET scheme can be employed. Specifically, a cutoff factor $\varepsilon$, $0 < \varepsilon < 1$, is used to terminate the iterative searching process early after $\mathcal{I}_{\text{e}} = \varepsilon \mathcal{I}_{\text{UB}}$ iterations in which no better solution is found. As a result, the number of searching iterations in the TS algorithm with ET is
	\begin{align*}
		\mathcal{I} = \min \left\{ \mathcal{I}_0 + \mathcal{I}_{\text{e}}, \mathcal{I}_{\text{UB}} \right\}, \numberthis \label{n_iter}
	\end{align*}
	where $\mathcal{I}_0$ is the number of iterations after which no better solution is found. The TS algorithm without ET requires $\mathcal{I}_{\text{UB}}$ searching iterations, which is also the upper bound on the number of iterations required when ET is applied. From \eqref{n_iter}, the following remarks are noted.
	
	\begin{remark}
		If ET occurs, the best solution found after $\mathcal{I}_0$ iterations and that found after $\mathcal{I} = \mathcal{I}_0 + \mathcal{I}_{\text{e}}$ iterations are the same, which is the final solution of the TS algorithm, i.e., $\hat{\vs}_{TS}$. Therefore, the earlier $\hat{\vs}_{TS}$ is found, the smaller $\mathcal{I}_0$ is required. This objective can be achieved by starting the moves in the TS algorithm with a good initial solution. Furthermore, if the initial solution is well taken such that it is likely to be a near-optimal solution, then no further searching iteration is required. In this case, the complexity of the TS algorithm becomes only that involved in finding the initial solution.
	\end{remark}
	
	\begin{remark}
		Another approach to find the optimal solution $\hat{\vs}_{TS}$ earlier and reduce $\mathcal{I}_0$ is to make efficient moves during the searching iterations in the TS algorithm. In other words, the moves can be guided so that $\hat{\vs}_{TS}$ is reached earlier. 
	\end{remark}
	
	\begin{remark}
		With the conventional ET criterion, $\mathcal{I}_e$ is fixed to $\mathcal{I}_e = \varepsilon \mathcal{I}_{\text{UB}}$. In large MIMO systems, a large $\mathcal{I}_{\text{UB}}$ is required to guarantee that the optimal solution is found, which results in a large $\mathcal{I}_e$. However, we note that once ET occurs, the further search over $\mathcal{I}_e$ iterations do not result in any performance improvement while causing significant computational burden for the TS algorithm. For example, in a $32 \times 32$ MIMO system with QPSK, $\mathcal{I}_{\text{UB}} = 800$ should be used to approximately achieve the performance of the SD scheme \cite{nguyen2019qr}. For $\varepsilon = 0.25$, $\mathcal{I}_e = 0.25 \times 800 = 200$ iterations are required before the termination, whereas $\hat{\vs}_{TS}$ was already found in the $(\mathcal{I}_0 = \mathcal{I} - 200)$th iteration. Therefore, a more efficient ET criterion is required to reduce the complexity of the TS algorithm with ET.
	\end{remark}
	
	Remarks 1--3 motivate us to propose a TS-based detection algorithm for complexity reduction with three design objectives: taking a good initial solution, using efficient moves in searching iterations so that $\hat{\vs}_{TS}$ is reached as soon as possible, and terminating the TS algorithm early based on an efficient ET criterion. In the next subsection, we propose the DL-aided TS algorithm with the application of DL to the TS detection for those objectives.

	\subsection{Proposed DL-aided TS algorithm}
	
	The main ideas of the proposed DL-aided TS algorithm can be explained as follows.
	
	\subsubsection{DL-aided initial solution}
	
	Unlike the conventional TS algorithms, in which the ZF, MMSE, or OSIC solution is taken as the initial solution \cite{zhao2007tabu}, the DL-aided TS algorithm employs a DNN to generate the initial solution.
	
	In this scheme, the most important task is to choose a DNN architecture that is not only able to approximate the transmitted signal vector with high accuracy, but also has low computational complexity. As discussed in Section II-B, a basic FC-DNN cannot achieve high BER performance for varying channels. By contrast, among the DetNet, ScNet, and FS-Net, the FS-Net requires the lowest complexity while achieving the best performance, which is demonstrated in Section \ref{sec_sim_result}. Therefore, we propose employing the FS-Net to find the initial solution in the DL-aided TS algorithm. Specifically, $\hat{\vs}$ obtained in step 6 of Algorithm \ref{al_dscnet} is taken as the initial solution of the DL-aided TS algorithm. As a result, the required complexity for this initialization phase is given in \eqref{c_dsn}.
	
	\subsubsection{Efficient moves in searching iterations}
	\label{sec_move}
	In the TS algorithm with ET, the complexity can be reduced if efficient moves are made during the iterative searching process, so that $\hat{\vs}_{TS}$ is found earlier, as discussed in Remark 2. For this purpose, we propose exploiting the difference between $\hat{\vs}^{[L]}$ and $\hat{\vs}$, which are the output of the last layer and the final solution of the FS-Net, respectively. We recall that $\hat{\vs}^{[L]} \in \setR^{N \times 1}$ can contain elements both inside and outside the alphabet $\setA$, as observed from step 4 in Algorithm \ref{al_dscnet} and Fig. \ref{fig_phi}. By contrast, we have $\hat{\vs} = \mathcal{Q} \left( \hat{\vs}^{[L]} \right) \in \setA^N$. 
	
	Let $\ve$ denote the distance between the elements of $\hat{\vs}$ and $\hat{\vs}^{[L]}$, i.e.,
	\begin{align*}
		\ve = \abs{ \hat{\vs}^{[L]} - \hat{\vs} } = \left[e_1, e_2, \ldots, e_N\right]^T.
	\end{align*}
	For QAM signals, the distance between two neighboring real symbols is two. Furthermore, from Fig. \ref{fig_phi}, we have $\hat{s}^{[L]} \in [-1, 1]$ for QPSK and $\hat{s}^{[L]} \in [-3, 3]$ for 16-QAM. Therefore, we have $0 \leq e_n \leq 1, n=1,2,\ldots,N$. It is observed that if $e_n \approx 0$, there is a high probability that the $n$th symbol in $\hat{\vs}$ is correctly approximated by the FS-Net, i.e., $s_n = \hat{s}_n$. By contrast, if $e_n \approx 1$, there is a high probability that $\hat{s}_n$ is an erroneous estimate, i.e., $s_n \neq \hat{s}_n$. Therefore, by examining the elements of $\ve$, we can determine the elements of $\hat{\vs}$ with high probabilities of errors.
	
	\textit{Example 1:} Consider a MIMO system with $N=2N_t=8$, QPSK, and
	\begin{align*}
		\hat{\vs}^{[L]} &= [0.1, -0.9, -0.2, 1, 0.25, 0.9, -1, 1]^T, \\
		\hat{\vs} &= [1,-1,-1,1,1,1,-1,1]^T.
	\end{align*}
	Then, we have
	\begin{align*}
		\ve = \abs{ \hat{\vs}^{[L]} - \hat{\vs} } = [0.9, 0.1, 0.8, 0, 0.75, 0.1, 0, 0]^T,
	\end{align*}
	which implies that $\hat{s}_1, \hat{s}_3$ and $\hat{s}_5$ can be incorrect with high probabilities. 
	
	In the following analysis, we refer to the symbols of $\hat{\vs}$ with high probabilities of being incorrect as the \textit{predicted incorrect symbols}. Furthermore, the $n$th element in $\hat{\vs}$ is determined as a predicted incorrect symbol if $e_n$ is beyond a predefined error-threshold $\gamma$, i.e., if $e_n > \gamma$. The error-threshold should be inversely proportional to the SNR, which leads to $\gamma = \frac{\lambda}{SNR}$, where $\lambda$ is chosen such that $\gamma \leq 1$. Furthermore, because $0 \leq e_n \leq 1, n=1,2,\ldots,N$, we set $\gamma$ to
	\begin{align*}
		\gamma = \min \left\{ \frac{\lambda}{SNR}, 0.5 \right\}. \numberthis \label{gamma}
	\end{align*}
	
	Let $n_e$ denote the number of predicted incorrect symbols in $\hat{\vs}$ ($n_e = 3$ in \textit{Example }1), and let $\mathbb{S}$ be the set of all candidates obtained by correcting $\hat{\vs}$. Then we have
	\begin{align*}
		\mathbb{S} = \mathcal{S}^{(1)} \cup \mathcal{S}^{(2)} \cup \ldots \cup \mathcal{S}^{(n_e)},
	\end{align*}
	where $\mathcal{S}^{(k)}, 1 \leq k \leq n_e,$ is a subset of $\mathbb{S}$ obtained by correcting $k$ elements of $\hat{\vs}$. The number of candidates in $\mathcal{S}^{(k)}$ is given as
	\begin{align*}
		\abs{\mathcal{S}^{(k)}} &= C_k^{n_e} \times (Q-1)^k  = \frac{n_e!(Q-1)^k}{k! (n_e-k)!},
	\end{align*}
	where $C_k^{n_e} = \frac{n_e!}{k! (n_e-k)!}$ is the number of combinations for choosing $k$ out of $n_e$ predicted incorrect symbols, and $Q$ is the number of real symbols in $\setA$ with $Q = \{2, 4, 8\}$ for QPSK, 16-QAM, and 64-QAM, respectively. Consequently, the total number of possible corrected vectors in $\mathbb{S}$ is 
	\begin{align*}
		\abs{\mathbb{S}} = \sum_{k=1}^{n_e} \frac{n_e!(Q-1)^k}{k! (n_e-k)!}.
	\end{align*}
	Now, denoting $\bar{\vs}^{\star}$ as the best vector obtained by correcting $\hat{\vs}$, we have
	\begin{align*}
		\bar{\vs}^{\star} = \arg \min_{\bar{\vs} \in \mathbb{S}} \norm{\vy - \mH \bar{\vs}}^2. \numberthis \label{s_tilde}
	\end{align*}
	In the case of large $n_e$ and high-order QAM schemes, $\abs{\mathbb{S}}$ becomes large; hence, high complexity is required to find $\bar{\vs}^{\star}$ in \eqref{s_tilde}.
	
	However, we note that correcting a symbol in $\hat{\vs}$ is equivalent to a move from $\hat{\vs}$ to one of its neighboring vectors in an iteration of the TS algorithm. Specifically, if $\hat{s}_k$ is likely to be wrong, then a move from $\hat{\vs}$ to its neighbor $\vx$ for the difference position of $k$ should be made. In this case, $\hat{s}_k$ and $x_k$ are neighboring symbols, and $\hat{s}_i = x_i$ for $i \neq k$. In this manner, many of the $n_e$ predicted incorrect symbols can be corrected after $t \geq n_e$ searching iterations of the TS algorithm with high probabilities. Let $\bar{\vs}^{\star}_{TS}$ be the solution found by correcting the predicted incorrect symbols of $\hat{\vs}$ after $t$ iterations. We have
		\begin{align*}
		\bar{\vs}^{\star}_{TS} = \arg \min_{\bar{\vs} \in \bar{\mathbb{S}}} \phi (\bar{\vs}), \numberthis \label{s_tilde_TS}
		\end{align*}
		where $\phi (\vx)$ is given in \eqref{phi_x}, and $\bar{\mathbb{S}} = \left\{ \bar{\vs}_{\{ 1 \}}, \ldots, \bar{\vs}_{\{ t \}} \right\} \subset \mathbb{S} $, with $\bar{\vs}_{\{ i \}}$ being the current candidate in the $i$th iteration, $i \leq t$.
	
	When $n_e$ is sufficiently small, the complexity involved in finding $\bar{\vs}^{\star}_{TS}$ with \eqref{s_tilde_TS} becomes much smaller than that for finding $\bar{\vs}^{\star}$ with \eqref{s_tilde}, as well as that for the conventional TS algorithm with \eqref{best_nb}. This is because $\bar{\mathbb{S}}$ is a subset of $\mathbb{S}$, and hence, unlike the conventional TS algorithm, in the $i$th iteration, $1 \leq i \leq t$, a reduced number of neighboring vectors are examined, which implies that the complexity required during $t$ iterations to find $\bar{\vs}^{\star}_{TS}$ can be low. If the incorrect symbols are predicted with high accuracy, there is a high probability that $\bar{\vs}^{\star}_{TS}$ is close to $\hat{\vs}_{TS}$. Therefore, only a small number of further iterations are required to reach $\hat{\vs}_{TS}$, and the total complexity of the TS algorithm can be reduced.

	\subsubsection{Proposed ET criteria}
	\label{sec_ET_criteria}
	In the conventional TS algorithm with ET, the algorithm is terminated early after $\mathcal{I}$ iterations as given in \eqref{n_iter}, where $\mathcal{I}_e$ is set to $\mathcal{I}_e = \varepsilon \mathcal{I}_{\text {UB}}$ with a fixed $\varepsilon$. Remark 3 shows that this stopping criterion is inefficient. Specifically, in many cases, the further searches over $\mathcal{I}_e$ iterations does not result in performance improvement while creating unnecessary complexity for the TS algorithm.
	
	In this work, to minimize the number of redundant searching iterations, we propose an adaptive ET criterion based on $n_e$. We have the following notes: 
	\begin{itemize}
		\item If $n_e = 0$, there is no predicted incorrect symbol in $\hat{\vs}$, which implies a high probability that $\hat{\vs}$ is already the optimal solution. In this case, no further searching iterations are needed, i.e., $\mathcal{I} = 0$.
		
		\item If $n_e \neq 0$ is small, there is a high probability that only a few elements of $\hat{\vs}$ are incorrect, which can be corrected after a small number of moves in the TS algorithm. Therefore, in this case, only a small $\mathcal{I}$ is required.
		
		\item In the case that $n_e \gg 1$, a sufficiently large number of searching iterations should be performed to guarantee the optimal performance.
	\end{itemize}
	
	Therefore, we propose using an adaptive cutoff factor $\hat{\varepsilon}$ depending on $\frac{n_e}{N}$ as follows:
	\begin{align*}
		\hat{\varepsilon} = \min \left\{ \varepsilon, \mu \frac{n_e}{N} \right\},
		\numberthis \label{cutoff}
	\end{align*}
	where $\mu$ is optimized through simulations to guarantee that a sufficient number of iterations is used in the DL-aided TS algorithm. Consequently, in the DL-aided TS algorithm, only $\hat{\mathcal{I}}_e = \hat{\varepsilon} \mathcal{I}_{\text{UB}}$ iterations are required before the searching process is terminated, with $\hat{\mathcal{I}}_e \leq {\mathcal{I}}_e$ because $\hat{\varepsilon} \leq \varepsilon$. From \eqref{cutoff}, we have $\hat{\mathcal{I}}_e \rightarrow 0$ as $n_e \rightarrow 0$, and $\hat{\mathcal{I}}_e \rightarrow \mathcal{I}_e = \varepsilon \mathcal{I}_{\text{UB}}$ as $n_e \rightarrow N$, implying the use of a smaller number of iterations for a smaller $n_e$, and vice versa.
	
	The DNN-aided TS algorithm is presented in Algorithm \ref{al_dnn_TS}. In step 1, $\hat{\vs}$ and $\hat{\vs}^{[L]}$ are obtained by the FS-Net scheme in Algorithm \ref{al_dscnet}, allowing $\ve$ to be computed in step 2 as the difference between $\hat{\vs}$ and $\hat{\vs}^{[L]}$. In step 4, the list $\mathcal{P}$ including the positions of the predicted incorrect elements of $\hat{\vs}$ is found based on $\gamma$, which is set in step 3. Then, $n_e$ is set to the size of $\mathcal{P}$ in step 5. The DL-aided TS algorithm is initialized in steps 6--11. Specifically, step 6 assigns $\hat{\vs}$ to the current candidate $\vc$ and pushes it to the tabu list. Then, the best solution $\hat{\vs}_{TS}$ and its metric $\phi \nt{\hat{\vs}_{TS}}$ are initialized as $\vc$ and $\phi (\vc)$, respectively, in step 7. The adaptive cutoff factor $\hat{\varepsilon}$ and $\hat{\mathcal{I}}_e$ are computed in steps 8 and 9, respectively. 
	
	In steps 12--30, $\hat{\vs}_{TS}$ is searched over $\mathcal{I}$ iterations. The first $t$ iterations are used to correct the predicted incorrect symbols in $\hat{\vs}$. In particular, in the $i$th iteration, $1 \leq i \leq t$, the best candidate $\vxb$ is found in step 15. In contrast, in the remaining $\mathcal{I} - t$ iterations, the conventional searching manner is used, where all the neighbors in $\setN$ are examined to find $\vxb$, as in steps 17 and 18. Comparing 15 to 18, it is observed that the proposed searching approach based on the predicted incorrect symbols requires lower complexity than the conventional searching approach because $\mathcal{S}(\vc) \subset \setN$. In steps 21--26, if a better solution is found, $\hat{\vs}_{TS}$ is updated, and at the same time, $count$ is set to zero to allow further moves to find a better solution. Otherwise, $count$ increases by one until it reaches $\hat{\mathcal{I}}_e$. The following steps update the best solution and the tabu list, and step 31 concludes the final solution after the searching phase is finished.

	\begin{algorithm}[t]
		\caption{DL-aided TS detection}
		\label{al_dnn_TS}
	\begin{spacing}{1.1}
		\begin{algorithmic}[1]
			\REQUIRE $\mH$, $\vy$
			\ENSURE $\hat{\vs}_{TS}$
			\STATE {Find $\hat{\vs}$ and $\hat{\vs}^{[L]}$ based on Algorithm \ref{al_dscnet}.}
			\STATE {$\ve = \abs{ \hat{\vs} - \hat{\vs}^{[L]} } = \left[e_1, e_2, \ldots, e_N\right]^T$}
			\STATE $\gamma = \min \left\{ \lambda / SNR, 0.5 \right\}$.
			
			\STATE Find $\mathcal{P}$ including the positions of elements of $\ve$ being larger than $\gamma$.
			\STATE Set $n_e = \abs{\mathcal{P}}$.
			
			\STATE {Initialize $\vc=\hat{\vs}$ and push $\vc$ to the tabu list.} 
			\STATE {$\hat{\vs}_{TS} = \vc, \phi \nt{\hat{\vs}_{TS}} = \phi \nt{\vc}$}

			\STATE Compute $\hat{\varepsilon}$ based on \eqref{cutoff}.
			\STATE $\hat{\mathcal{I}}_e = \hat{\varepsilon} \mathcal{I}_{\text{UB}}$.
			
			\STATE {$i = 1, count = 0$}
			\STATE $\mathcal{S}^{(0)} = \vc$
			
			\WHILE {$i \leq \mathcal{I}_{\text{UB}}$ \textbf{and} $count \leq \hat{\mathcal{I}}_e$}
			
			\IF {$i \leq t$}
			\STATE Find $\mathcal{S} (\vc)$ which includes only the neighbors of $\vc$ with the difference positions in $\mathcal{P}$. 
			\STATE $\vxb = \arg \min_{\vx \in \mathcal{S}  (\vc)} \{ \phi (\vx) \}$
			\ELSE
			\STATE {Find the neighbor set $\setN$ of $\vc$.}
			\STATE {$\vxb = \arg \min_{\vx \in \setN } \{ \phi (\vx) \}$}
			\ENDIF
			
			\STATE {Update the current candidate: $\vc = \vxb$}
			\IF {$\phi \nt{\vc} < \phi \nt{\hat{\vs}_{TS}}$}
			\STATE {Update the best solution: $\hat{\vs}_{TS} = \vc$.}
			\STATE {$count  = 0$}
			\ELSE
			\STATE {$count = count + 1$}
			\ENDIF
			
			\STATE {Release the first element in the tabu list if it is full.}
			\STATE {Push $\vc$ to the tabu list and update its length.}
			
			\STATE $i = i + 1$
			\ENDWHILE
			\STATE {The final solution is the best solution $\hat{\vs}_{TS}$ found so far.}
		\end{algorithmic}
	\end{spacing}
	\end{algorithm}

	\section{Simulation Results}
	\label{sec_sim_result}
	In this section, we numerically evaluate the BER performance and computational complexities of the proposed FS-Net detection architecture and the DL-aided TS algorithm. In our simulations, each channel coefficient is assumed to be an i.i.d. zero-mean complex Gaussian random variable with a variance of $1/2$ per dimension. The SNR is defined as the ratio of the average symbol power $\smt$ to the noise power $\smn$. 
	
	\subsection{Training DNNs}
	We follow the training model in \cite{samuel2019learning} and \cite{gao2018sparsely}. Specifically, we use the Adam optimizer \cite{kingma2014adam}, which is a variant of the stochastic gradient descent method \cite{rumelhart1988learning, bottou2010large} for optimizing the DNNs. The DNNs are implemented by using Python with the Tensorflow library \cite{abadi2016tensorflow} and a standard Intel i7-6700 processor. For the training phase, a decaying learning rate of 0.97 and starting learning rate of 0.0001 are used. Furthermore, the number of layers in the DNNs is set to at least 10 and varies depending on the sizes of the considered MIMO systems. We train the DNNs for 20000 iterations with a batch size of 2000 samples. For each sample, $\vs, \mH$, and $\vy$ are independently generated from \eqref{real SM}, then $\mH^T \mH$ and $\mH^T \vy$ are computed accordingly. 
	
	\begin{figure}[t]
		\centering
		\includegraphics[scale=0.62]{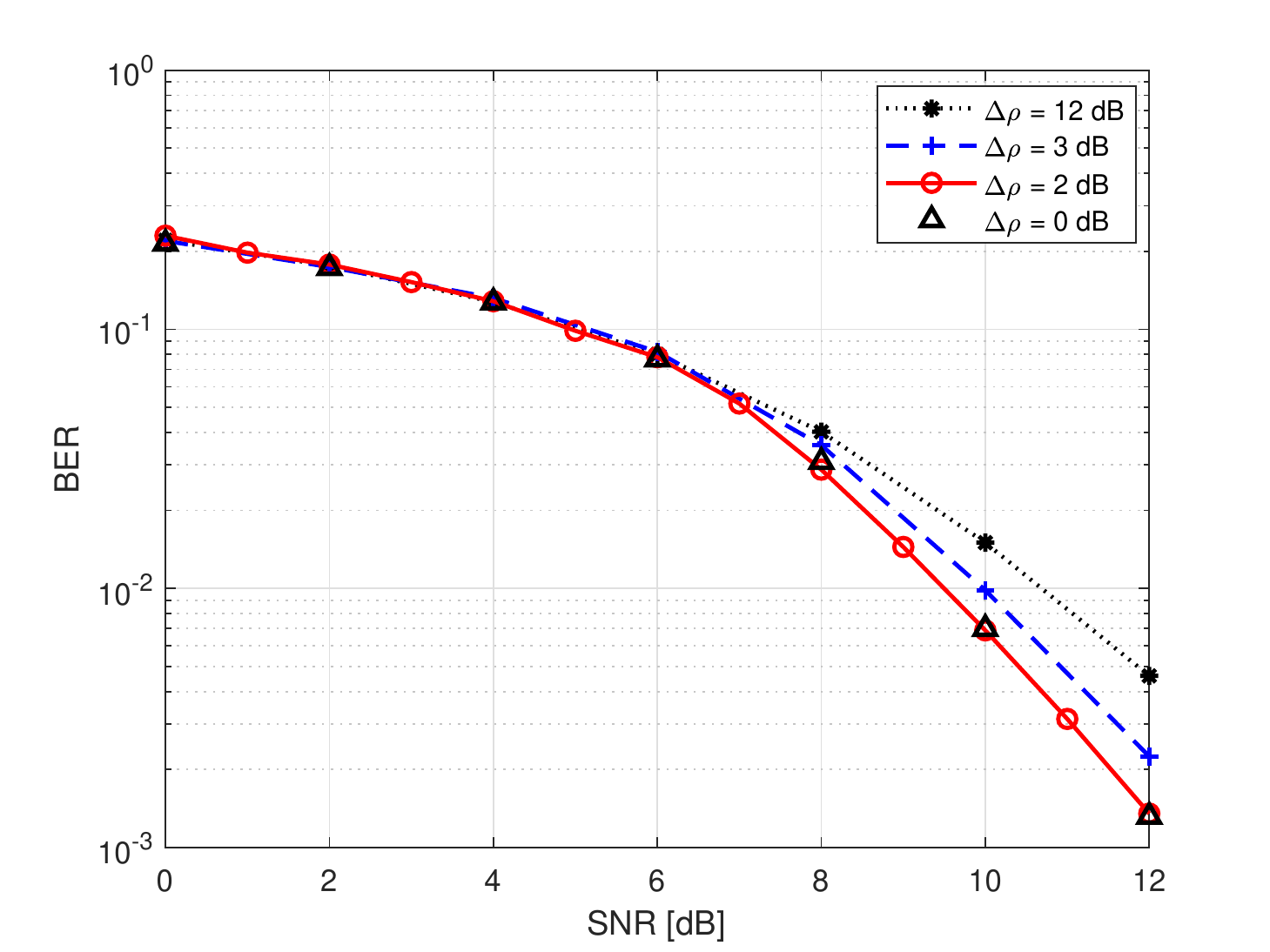}
		\caption{BER performance of the proposed FS-Net architecture for various training SNR ranges corresponding to $\Delta \rho = \left\{ 0, 2, 3, 12 \right\}$ dB for $(N_t \times N_r) = (32 \times 32)$,  $L = 15$ with QPSK.}
		\label{fig_ber_testing_FSNet}
	\end{figure}

	In the training phase of the DetNet and ScNet, the noise variance can be randomly generated so that the SNR is uniformly distributed in the range $[\rho_{min}, \rho_{max}]$. In this training method, the network needs to adapt to a wide range of SNR. Therefore, a large number of layers and neurons need to be employed, which requires high computational complexity and a long training time. To resolve this problem, we consider training multiple networks for multiple divided SNR ranges. The only disadvantage of this method is its requirement of additional memory to save multiple trained models. However, the network can learn faster to achieve higher accuracy.
	
	To show the performance improvement by reducing the SNR range in Fig. \ref{fig_ber_testing_FSNet}, we train and test the FS-Net for $32 \times 32$ MIMO with QPSK and $\Delta \rho = \left\{ 0, 2, 3, 12 \right\}$ dB with $\alpha = 1$ and $\beta = 0.5$ being set for the loss function in \eqref{loss_dscnet}. Here, $\Delta \rho = \rho_{max} - \rho_{min}$ represents the SNR interval for which a DNN is trained. For example, $\Delta \rho = 0$ dB means that each FS-Net is trained only for a particular SNR. In contrast, for $\Delta \rho = 12$ dB, it is trained for the entire considered SNR range as the DetNet and ScNet are trained in \cite{samuel2017deep, samuel2019learning, gao2018sparsely}. In each training epoch, the SNR is randomly selected in the corresponding range. For performance evaluation, an appropriately trained FS-Net is selected for each considered SNR. For example, to test the FS-Net at SNR $= 10$ and $11$ dB, the FS-Net trained for the SNR range of $[9, 11]$ dB is employed.
	
	The BER performance for $\Delta \rho = \left\{ 0, 2, 3, 12 \right\}$ dB is shown in Fig. \ref{fig_ber_testing_FSNet}. It is observed that as $\Delta \rho$ decreases, the BER performance is significantly improved. In particular, the FS-Nets with $\Delta \rho = 0$ and $2$ dB achieve the best BER performance. For realistic environments with arbitrary SNRs, the DNN trained only for a specific SNR, i.e., $\Delta\rho = 0$ dB, can be impractical; therefore, for the remaining simulations, we train the DetNet, ScNet, and FS-Net with $\Delta \rho = 2$ dB. We note that for different SNR ranges, the trained DNNs have different weights and biases but the same network architecture. Therefore, they do not result in any further computational complexity in the operations of the FS-Net and DL-aided TS algorithm.

	\subsection{Performance and complexity of the proposed FS-Net architecture}
	
	\begin{figure}[t]
		\subfigure[$(N_t \times N_r) = (32 \times 64)$,  $L = 15$]
		{
			\includegraphics[scale=0.62]{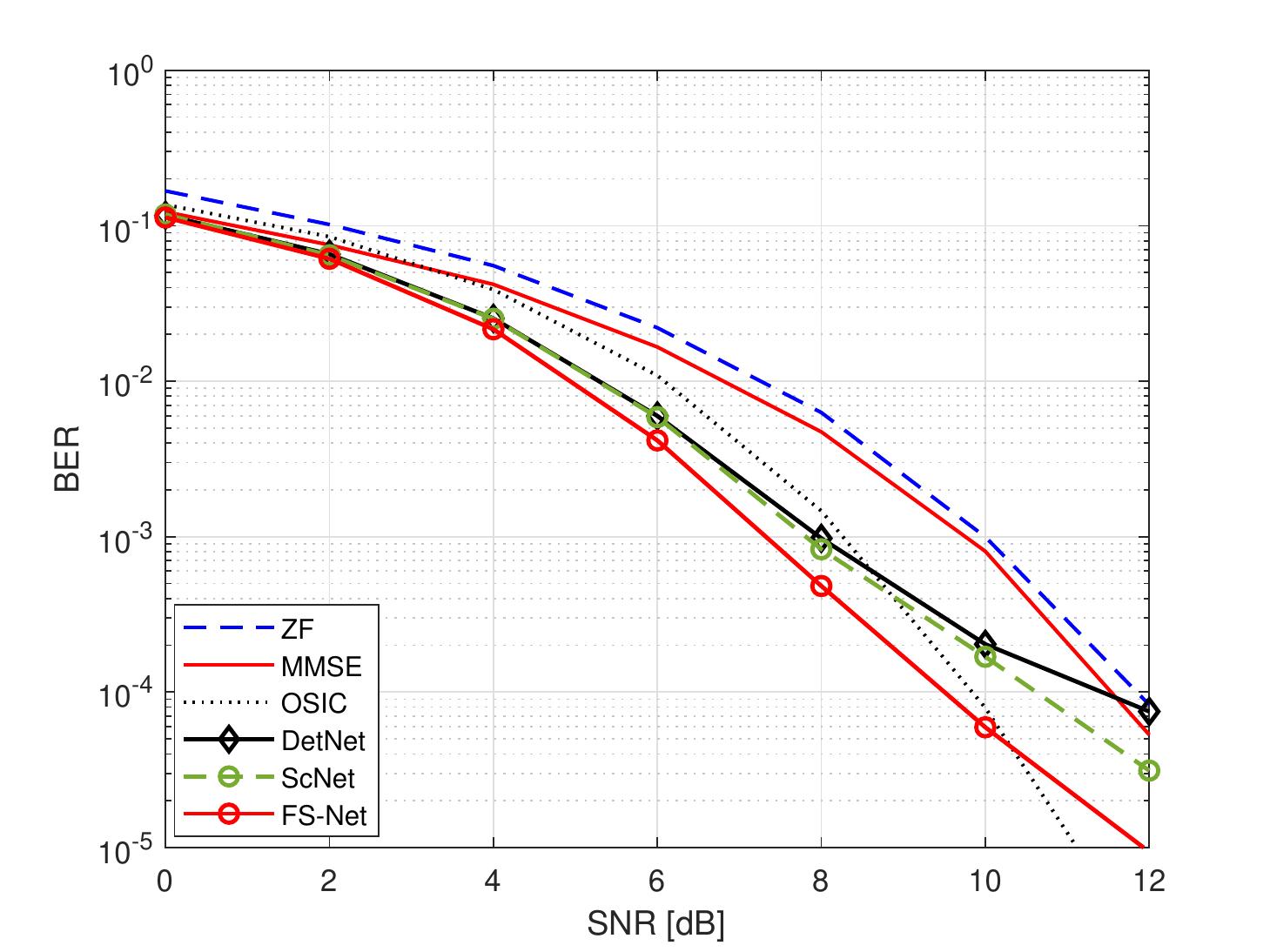}
			\label{fig_ber_DSN_Nt_smaller_than_Nr}
		}
		\subfigure[$(N_t \times N_r) = (32 \times 32)$,  $L = 15$]
		{
			\includegraphics[scale=0.62]{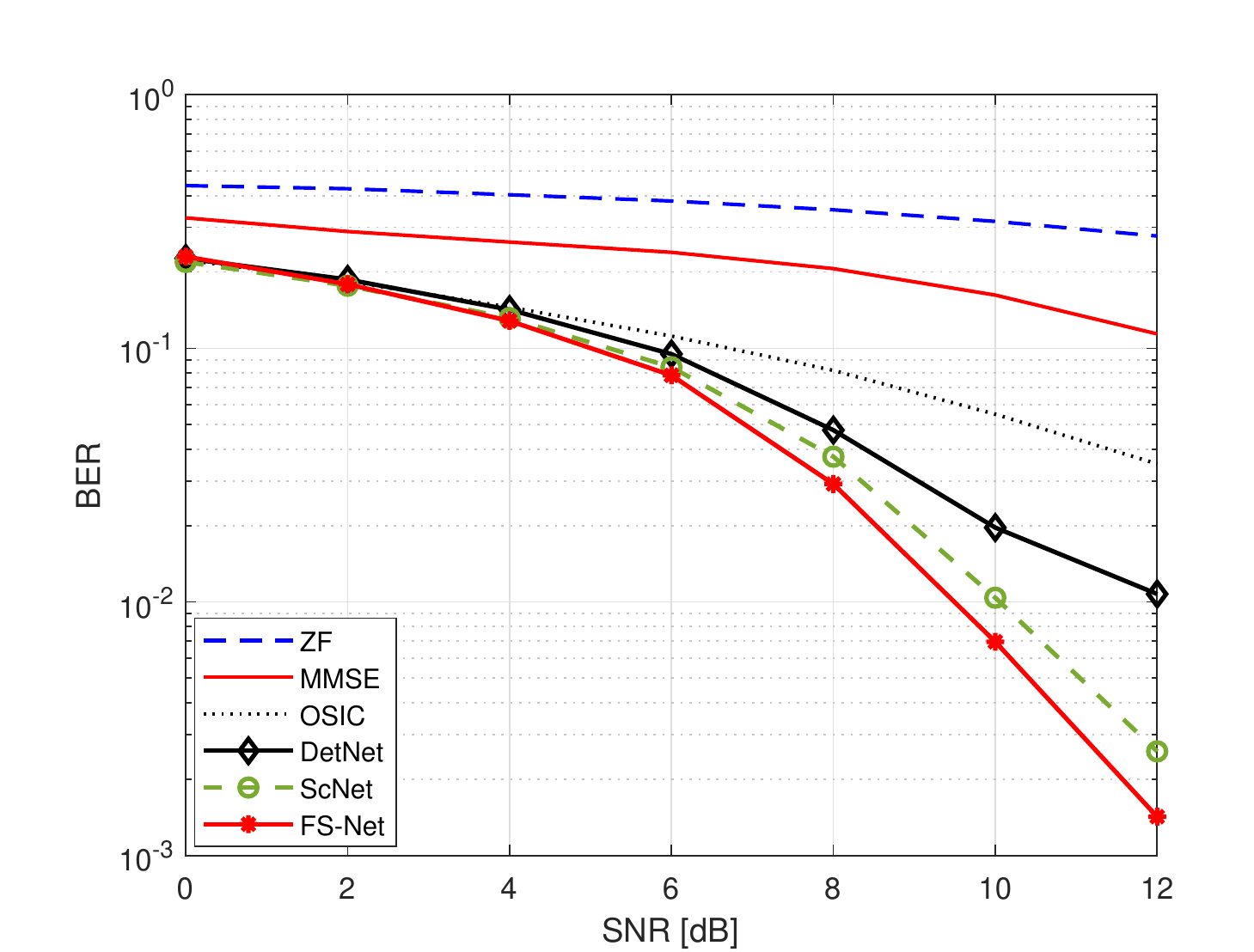}
			\label{fig_ber_DSN_Nt_equal_Nr}
		}
		\caption{BER performance of the proposed FS-Net architecture in comparison with those of the DetNet, ScNet, ZF, MMSE, OSIC schemes with  QPSK.}
		\label{fig_ber_DSN}
	\end{figure}
	
	In Fig. \ref{fig_ber_DSN}, we show the BER performance of the proposed FS-Net in comparison to those of the DetNet \cite{samuel2019learning} and ScNet \cite{gao2018sparsely} for two scenarios: $N_t < N_r$ and $N_t = N_r$. Specifically, we assume $(N_t \times N_r) = (32 \times 64)$ in Fig. \ref{fig_ber_DSN_Nt_smaller_than_Nr} and $(N_t \times N_r) = (32 \times 32)$ in Fig. \ref{fig_ber_DSN_Nt_equal_Nr}, and $L=15$ and QPSK are assumed in both scenarios. For comparison, linear ZF/MMSE and ZF-based OSIC schemes \cite{paulraj2003introduction} are also considered. From Fig. \ref{fig_ber_DSN}, the following observations are noted:
	\begin{itemize}
		\item For both considered scenarios, the DNN-based schemes outperform the linear ZF and MMSE receivers. The OSIC receiver only performs better than the DetNet and ScNet schemes for SNR $> 8$ dB and the FS-Net for SNR $> 10$ dB in the $32 \times 64$ MIMO system, whereas in the $32 \times 32$ MIMO system, it is outperformed by the DNN-based schemes.
		
		\item It is clear that among the compared schemes, the proposed FS-Net scheme achieves the best performance in both considered scenarios. Specifically, in both figures, it achieves approximately the SNR gains of 1 dB with respect to the ScNet and more than 2 dB with respect to the DetNet architecture in the high-SNR region.
	\end{itemize}
	
	\begin{table}[t]
		\renewcommand{\arraystretch}{1.3}
		\caption{Computational complexities of the DetNet, ScNet, and FS-Net for $32 \times 64$ and $32 \times 32$ MIMO with QPSK and $L=15$.}
		\label{tab_comp_DSN}
		\centering
		\begin{tabular}{c|c|c|c}
			\hline
			Number of operations   &  DetNet & ScNet & FS-Net \\
			\hline
			\hline
			
			\makecell{$32 \times 64$ MIMO}  & $2.168 \times 10^7$ & $1.188 \times 10^6$  & $1.187 \times 10^6$ \\
			\hline
			
			\makecell{$32 \times 32$ MIMO}  & $1.627 \times 10^6$  & $6.476 \times 10^5$ & $6.467 \times 10^5$  \\
			\hline

		\end{tabular}
	\end{table}
	
	In Table \ref{tab_comp_DSN}, we show the complexities of the DetNet, ScNet, and FS-Net schemes. As in Fig. \ref{fig_ber_DSN}, we consider $(N_t \times N_r) = \left\{(32 \times 64),(32 \times 32)\right\}$ and $L=15$. The complexities of the DetNet, ScNet, and FS-Net are computed based on \eqref{c_detnet_qpsk}, \eqref{c_scnet}, and \eqref{c_dsn}, respectively. It is seen from Fig. \ref{fig_ber_DSN} and Table \ref{tab_comp_DSN} that the proposed FS-Net architecture not only achieves better BER performance, but also requires lower computational complexity than the prior DetNet and ScNet schemes. Specifically, the complexity of the FS-Net is much lower than that of the DetNet and slightly lower than that of the ScNet. Therefore, we can conclude that among the compared architectures, the proposed FS-Net achieves the best performance--complexity tradeoff.
	
	\subsection{Performance of the proposed DL-aided TS algorithm}
	
	\begin{figure}[t]
		\centering
		\includegraphics[scale=0.62]{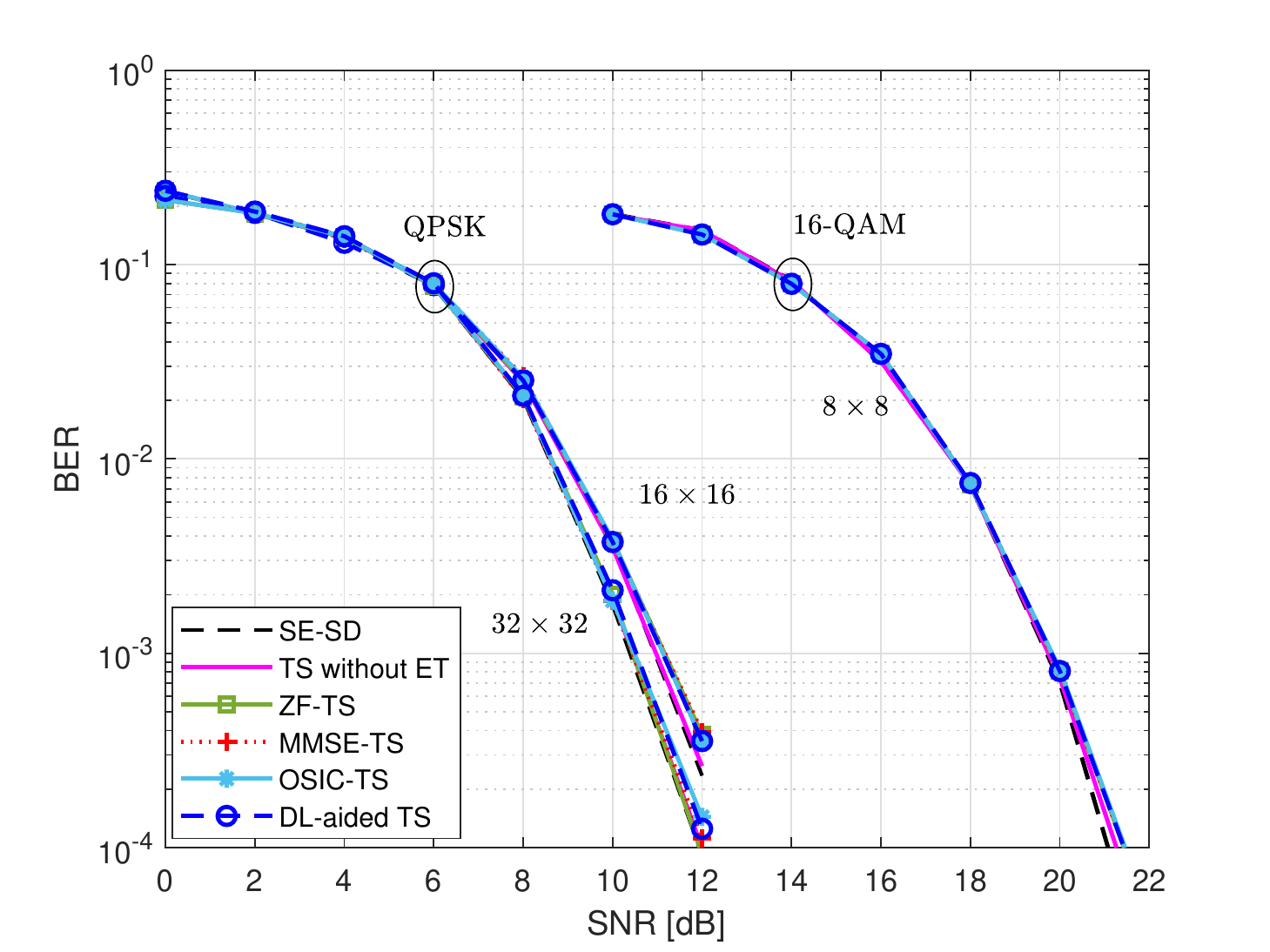}
		\caption{BER performance of the proposed DL-aided TS algorithm in comparison with those of the ZF-TS, MMSE-TS, OSIC-TS, TS without ET, and the SE-SD receivers for $(N_t \times N_r) = \left\{ (16 \times 16), (32 \times 32)\right\}$,  $L = \left\{10, 15\right\}$ with QPSK and $(N_t \times N_r) = (8 \times 8)$,  $L = 15$ with 16-QAM.}
		\label{fig_ber_QPSK}
	\end{figure}
	
	In this section, we show the BER performance of the proposed DL-aided TS algorithm, which is compared to those of the Schnorr--Euchner SD (SE-SD) scheme \cite{agrell2002closest} and the conventional TS algorithm where ET is not applied \cite{zhao2007tabu}, and the ZF solution is used as the initial solution. Furthermore, to show the advantage of the proposed DL-aided TS algorithm, it is also compared to the TS schemes with ET and ZF (ZF-TS), MMSE (MMSE-TS), and OSIC (OSIC-TS) for initial solutions.
	
	\begin{table}[t]
		\renewcommand{\arraystretch}{1.3}
		\caption{Simulation parameters used in Figs. \ref{fig_ber_QPSK}--\ref{fig_comp_8_QAM}}
		\label{tab_comp_DSN}
		\centering
		\begin{tabular}{c|c|c|c}
			\hline
			Parameters   &  $16 \times 16$, QPSK & $32 \times 32$, QPSK & $8 \times 8$, 16-QAM \\
			\hline
			\hline
			
			\makecell{$\mathcal{I}_{\text {UB}}$}  & $400$ & $800$  & $2500$ \\
			\hline
			
			\makecell{$\varepsilon$}  & $0.4$  & $0.4$ & $0.4$  \\
			\hline
			
			\makecell{$\lambda$}  & $1/32$  & $1/64$ & $12/16$  \\
			\hline
			
			\makecell{$\mu$}  & $5$  & $3$ & $0.5$  \\
			\hline
			
			\makecell{$L$}  & $10$  & $15$ & $20$  \\
			\hline
		\end{tabular}
	\end{table}
	
	For comparison, various systems are considered, including the $16 \times 16$ and $32 \times 32$ MIMO with QPSK and $8 \times 8$ MIMO with 16-QAM, and the simulation parameters are presented in Table II. Specifically, $\mathcal{I}_{\text {UB}}$ is set based on \cite{nguyen2019qr}, which guarantees that the TS schemes can approximately achieve the BER performance of the SE-SD receiver. The length of the tabu list is set to $P = \mathcal{I}_{\text {UB}}/2$ so that TS algorithms approximately approach the optimal performance \cite{zhao2007tabu, nguyen2019qr}. In all the considered systems, we use $\varepsilon = 0.4$, for which the employment of ET results in only marginal performance loss. Furthermore, for the DL-aided TS algorithm, the values of $\lambda$ and $\mu$ are optimized through simulations, and $t=4 n_e$ is assumed for the number of searching iterations in which only the predicted incorrect symbols are considered for neighbor search. Regarding the FS-Net architecture, the number of layers, $L$, should be chosen such that the FS-Net-based solution can be found with a good performance--complexity/training time trade-off. For this, the FS-Net for each MIMO systems is trained and tested with different values of $L$, and the chosen ones are presented in Table II. 
	
	It is observed from Fig. \ref{fig_ber_QPSK} that in all the considered systems, the TS algorithms with ET, namely, ZF-TS, MMSE-TS, OSIC-TS, and DL-aided TS, all have approximately the same BER performance. In particular, although ET is employed, only marginal performance losses are seen for the ZF-TS, MMSE-TS, OSIC-TS, and DL-aided TS schemes with respect to the conventional TS algorithm without ET. Furthermore, with the chosen parameters, all the considered TS algorithms approximately achieve the performance of the SE-SD scheme.
	
	\subsection{Complexity reduction of the DL-aided TS algorithm}
	
	\begin{figure}[t]
		\subfigure[$(N_t \times N_r) = (16 \times 16)$,  $L = 10$]
		{
			\includegraphics[scale=0.62]{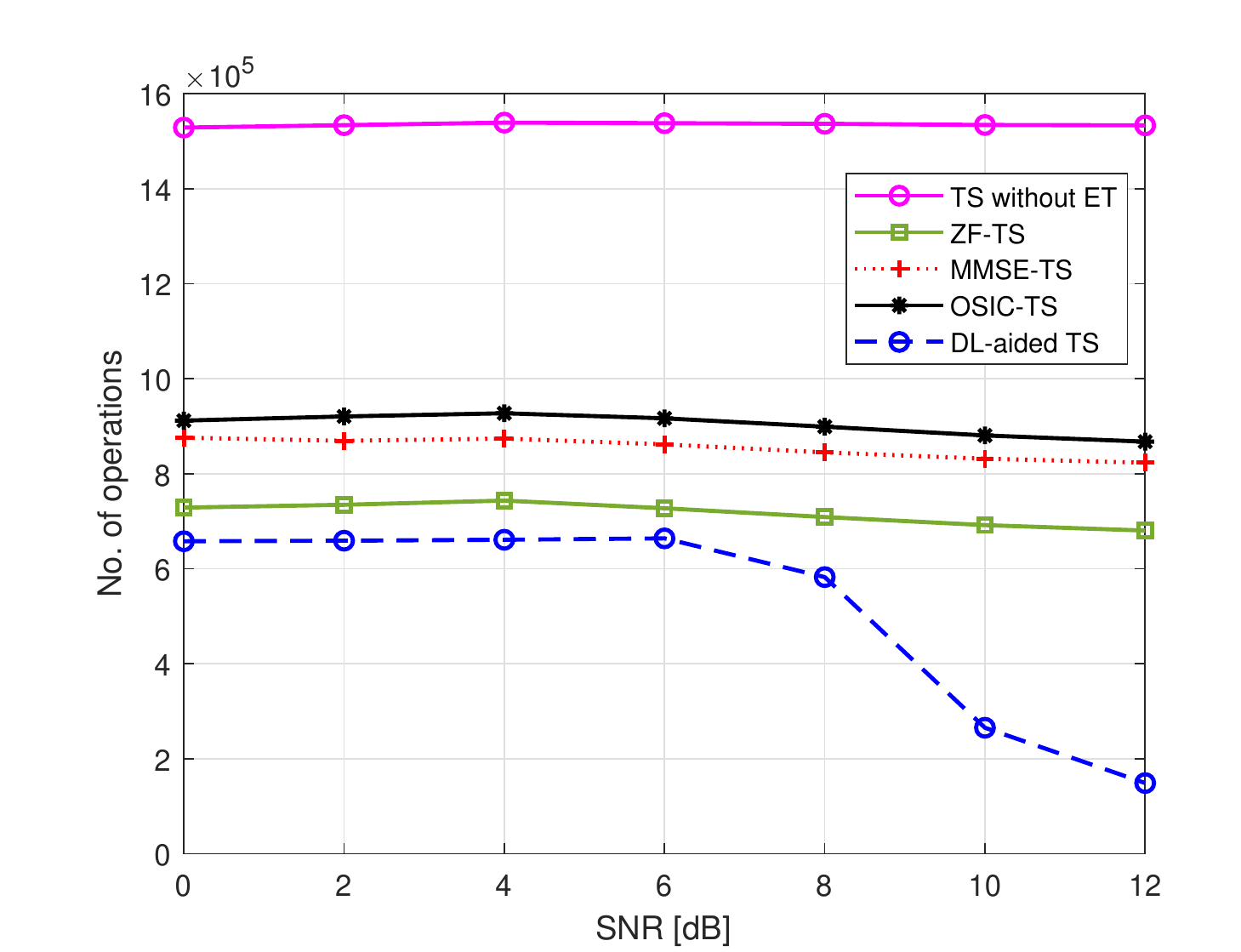}
			\label{fig_comp_16}
		}
		\subfigure[$(N_t \times N_r) = (32 \times 32)$,  $L = 15$]
		{
			\includegraphics[scale=0.62]{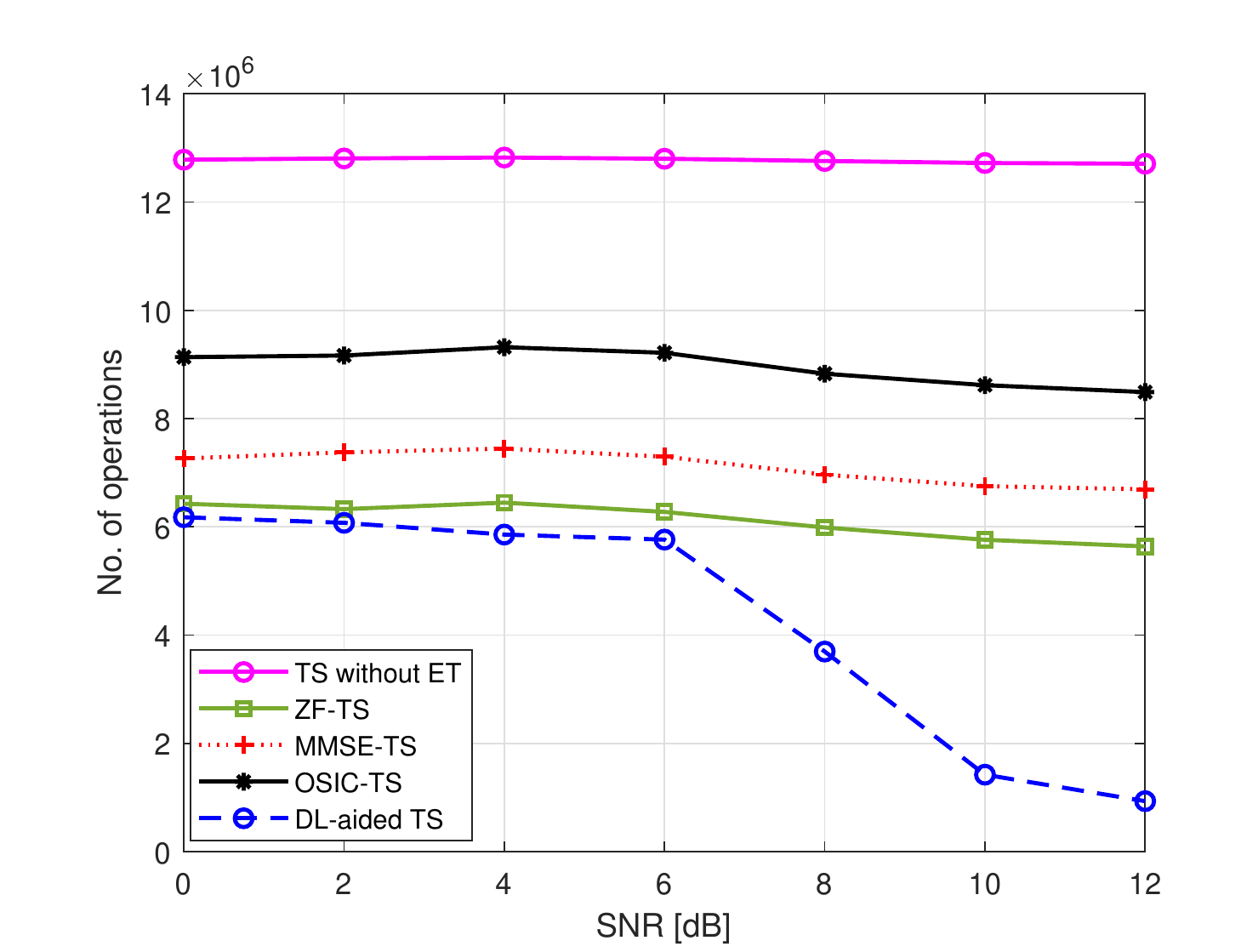}
			\label{fig_comp_32}
		}
		\caption{Complexity of the proposed DL-aided TS algorithm in comparison with those of the ZF-TS, MMSE-TS, OSIC-TS, and TS without ET with QPSK.}
		\label{fig_comp}
	\end{figure}
	
	\begin{figure}[t]
		\centering
		\includegraphics[scale=0.62]{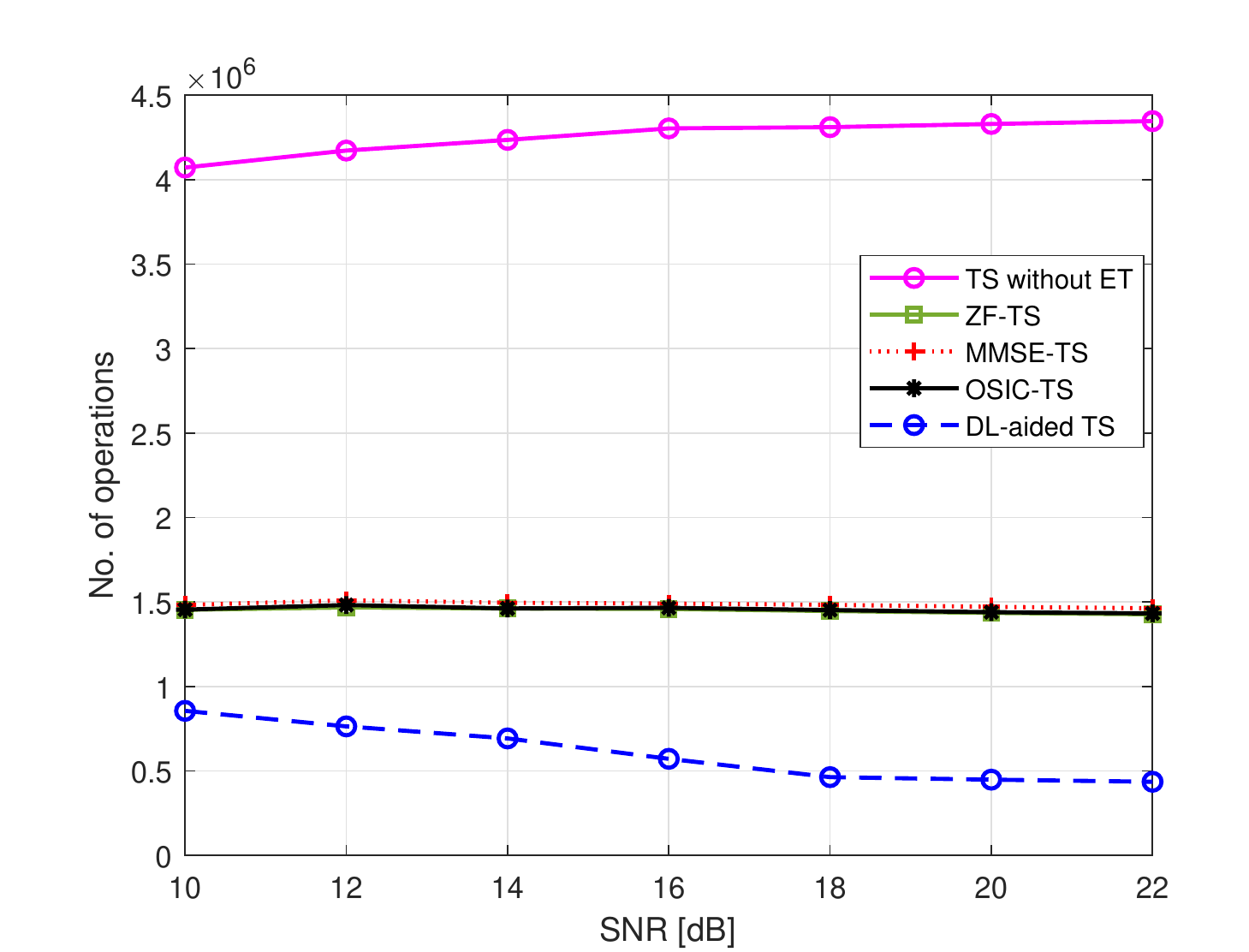}
		\caption{Complexity of the proposed DL-aided TS algorithm in comparison with those of the ZF-TS, MMSE-TS, OSIC-TS, and TS without ET with $(N_t \times N_r) = (8 \times 8)$,  $L = 20$, and 16-QAM.}
		\label{fig_comp_8_QAM}
	\end{figure}

	In Figs. \ref{fig_comp} and \ref{fig_comp_8_QAM}, the complexity of the DL-aided TS algorithm is compared to those of the conventional TS, ZF-TS, MMSE-TS, and OSIC-TS for $16 \times 16$ and $32 \times 32$ MIMO systems with QPSK and $8 \times 8$ MIMO with 16-QAM. To ensure that the compared algorithms have approximately the same BER performance, the simulation parameters are assumed to be exactly the same as those in Fig. \ref{fig_ber_QPSK} and are presented in Table II. It is clear from Figs. \ref{fig_comp} and \ref{fig_comp_8_QAM} that the TS algorithms with ET can significantly reduce the complexity of the TS algorithm, while resulting in only marginal performance losses, as discussed in the previous subsection. Furthermore, among the TS algorithms with ET, the DL-aided TS scheme has the lowest computational complexity. For example, in Fig. \ref{fig_comp_32}, at SNR $=12$ dB, the complexity of the DL-aided TS algorithm is only approximately $16.5\%$, $13.9\%$, $10.9\%$, and $7.3\%$ those of the ZF-TS, MMSE-TS, OSIC-TS, and conventional TS scheme, respectively. In Fig. \ref{fig_comp_8_QAM}, the complexity-reduction ratio of the proposed DL-aided TS algorithm is approximately $70\%$ with respect to the other ET-based TS algorithms and $82\%$ with respect to the conventional TS without ET at SNR $=22$ dB. Furthermore, we see that unlike the ZF-TS, MMSE-TS, OSIC-TS, and conventional TS schemes, the complexity of the proposed DL-aided TS algorithm decreases with SNR. The reason for that is as follows:
		\begin{itemize}
			\item At low SNRs, the FS-Net scheme in Algorithm \ref{al_dscnet} has low accuracy, which leads to large $n_e$. In this case, we have $\hat{\varepsilon} \approx \varepsilon$, and hence, the complexity reduction is only obtained by reduced search space during the first iterations. However, for large $n_e$, the reduction of search space becomes small, which leads to a relatively small gain in complexity reduction.
			
			\item At high SNRs, the FS-Net-based initial solution is generated with high accuracy and small $n_e$. Therefore, a significant complexity reduction is achieved with the DL-aided TS algorithm owing to a reduced cutoff factor $\hat{\varepsilon}$ as well as the reduced search space during the first $t$ iterations.
		\end{itemize}
	
	\section{Conclusion}
	In this study, we have presented the FS-Net detection scheme, which is a DNN-aided symbol-detection algorithm for MIMO systems. Unlike the prior DetNet and ScNet schemes, the input vector of the FS-Net is optimized such that a reduced number of connections is required in each layer. Furthermore, the correlation between the input and output signals is considered in the loss function of the FS-Net, allowing it to be trained better with faster convergence compared to DetNet and ScNet. The optimized input, network connection, and loss function lead to the reduced complexity and improved performance of the proposed FS-Net scheme.

	The proposed FS-Net is then incorporated into the TS algorithm. Specifically, it is used to generate the initial solution with enhanced accuracy for the TS algorithm. Furthermore, based on the initial solution given by the FS-Net, the iterative searching phase of the TS algorithm is improved by predicting incorrect symbols. This can facilitate more efficient moves in the TS algorithm so that the optimal solution is more likely to be reached earlier. We also propose using an adaptive ET criterion, in which the cutoff factor is adjusted based on the accuracy of the FS-Net-based initial solution. As a result, a small number of searching iterations is taken for a reliable initial solution, which leads to a reduction in the overall complexity of the TS algorithm. 
	
	Our simulation results show that the proposed FS-Net scheme not only outperforms the linear ZF/MMSE and the OSIC receivers but also achieves improved performance and reduced complexity with respect to the DetNet and ScNet. Furthermore, with almost the same BER performance, the proposed DL-aided TS scheme with the FS-Net-based initial solution requires much lower complexity than the other TS algorithms with ET, such as ZF-TS, MMSE-TS, and OSIC-TS. We note that the proposed DL-aided TS scheme can be combined with existing TS algorithms, such as LTS \cite{srinidhi2011layered}, R3TS \cite{datta2010random}, and QR-TS \cite{nguyen2019qr}, for better initialization and more efficient ET.

	\bibliographystyle{IEEEtran}
	\bibliography{IEEEabrv,Bibliography}

\end{document}